\documentclass[12pt]{article}

\usepackage{amsmath}
\usepackage{amssymb}
\usepackage[margin=3cm]{geometry}
\usepackage{graphicx}
\usepackage{cite}
\numberwithin{equation}{section}
\allowdisplaybreaks
\usepackage{hyperref}
\usepackage{chngcntr}
\counterwithin{table}{section}

\def\be{\begin{equation}}
\def\ee{\end{equation}}
\def\ba{\begin{eqnarray}}
\def\ea{\end{eqnarray}}

\begin{document}

\counterwithin{figure}{section}


\title{Effective field theory of a single scalar pion field for large scale structure in the Universe}
\author{Lara Celik\thanks{lcelik01@manhattan.edu}, Bart Horn\thanks{bhorn01@manhattan.edu}, Bhavya Mishra\thanks{bmishra01@manhattan.edu},
David Muqattash\thanks{dmuqattash01@manhattan.edu}}
\date{\normalsize\textit{Department of Mathematics and Physics, Kakos School of Arts and Sciences, \\ Manhattan University, New York, NY 10471, USA}}

\maketitle

\noindent \abstract{We discuss the effective field theory of large scale structure in terms of a single scalar degree of freedom, corresponding to the velocity potential of the matter fluid in a $\Lambda$CDM universe.  This cosmic ``pion'' field is nonlinearly related to the overdensity and the gravitational potential, and corresponds to the Goldstone boson of spontaneously broken spacetime symmetry, allowing us to organize perturbation theory in a systematic way while keeping the symmetries manifest.  We develop the effective field theory of the pion field to next-to-leading order, and we use it to calculate the corrections to the power spectrum and to check that these are consistent with the consistency relations of spontaneously broken spacetime symmetry.  We compare our results against computer simulations for the evolution of large scale structure in the pion field picture, and we make use of N-body simulations to measure EFT coefficients and analyze the growth of additional degrees of freedom in the deep nonlinear regime.  We conclude with a discussion of how the pion field picture may help suggest new variables for analyzing simulations and experimental surveys of large scale structure.}

\section{Introduction}

Over the next five to ten years, existing and upcoming surveys of large scale structure will map the cosmos to unprecedented levels of precision.  Despite great progress in constraining cosmic parameters to the per cent level using cosmic microwave background radiation (CMB) data, however, many unanswered questions remain.  Measuring or further constraining parameters such as the spatial curvature, primordial non-Gaussianity, the Hubble expansion parameter, early or evolving dark energy, or potential modifications to general relativity can give us a great deal of information about the primordial era and model building.  Large scale structure (LSS) probes the inhomogeneous Universe on similar length scales as the CMB; however, LSS maps the Universe in three dimensions, and over different eras in cosmic time.  Although there is potentially a great deal of data to be learned from, the data is more complex due to the nonlinear evolution, as well as galaxy and other tracer bias, redshift space distortions and other complications.  N-body techniques for simulating large scale structure, while extremely powerful and accurate, are computationally expensive on large scales.  Perturbative methods work well in the weakly non-linear regime but become inefficient as nonlinear interactions begin to become important.  In order to extract parameters of primordial physics from the late-time data, it is of interest to investigate new analytical and computational techniques.  

Effective field theory techniques inspired by high energy physics have been successfully applied to problem of cosmological clustering, principally through the EFT for LSS formalism introduced in \cite{Baumann:2010tm, Carrasco:2012cv} (see also \cite{Pietroni:2011iz, Carroll:2013oxa} for related work on coarse-graining of cosmological perturbation theory), which characterizes matter inhomogeneities as an effective fluid described by a density, velocity, sound speed, viscosity, and other parameters on large scales.  This formalism has been further developed and extended to include higher order loop effects and renormalization \cite{Carrasco:2013mua, Pajer:2013jj, Foreman:2015uva, Baldauf:2015aha, Foreman:2015lca}, velocity calculations \cite{Mercolli:2013bsa}, Lagrangian space formalism \cite{Porto:2013qua, Senatore:2014via, Vlah:2015sea}, baryonic effects \cite{Lewandowski:2014rca}, and redshift space formalism \cite{Lewandowski:2015ziq}, to give an illustrative but by no means exhaustive list.

In this paper we investigate the effective field theory of large scale structure using the pion effective field theory developed by Nicolis and Vernizzi\cite{Nicolis} and described in further detail in \cite{Horn:2014rta}.  Pressureless fluid in an expanding $\Lambda$CDM universe can be described in terms of a single scalar field which in the non-relativistic limit corresponds to the velocity potential 
\begin{equation}
\vec{v} = \vec{\nabla} \pi\,,
\end{equation}
and the Poisson, Navier-Stokes and Friedmann equations can be used to relate $\pi$ to more familiar variables such as the Newtonian potential $\Phi$ and the overdensity $\delta$.  The $\pi$ field has several advantages from a field theoretic point of view: it combines the variables $\vec{v}, \Phi, \delta$ into a single scalar degree of freedom (as long as vorticity can be neglected, which is the case on scales larger than the $\approx 10$ Mpc scale of nonlinear clustering), and as discussed in \cite{Horn:2014rta}, it is the Goldstone field corresponding to the breaking of certain diffeomorphism symmetries.  We therefore expect that the pion field is a natural description for organizing the effective field theory of large scale structure, and this formalism can be compared to the standard EFT for LSS formalism introduced in \cite{Baumann:2010tm, Carrasco:2012cv}.  Since the $\pi$ field is non-linearly related to more conventional variables, it may also help suggest new experimental signatures for extracting the parameters of primordial physics.  In particular, we expect that the pion field is a natural framework for expressing the realization of consistency relations in large scale structure on both subhorizon and superhorizon scales \cite{Kehagias:2013yd, Peloso:2013spa, Creminelli:2013mca, Horn:2014rta} (see also \cite{Esposito:2019jkb, Goldstein:2022hgr} for studies applying these consistency conditions to simulations and data).

This paper is organized as follows.  In \S2 we review the pion field description of large scale structure, first by starting with the Navier-Stokes equations for pressureless fluid in expanding spacetime and then by deriving the effective field theory of the pion field systematically from a fully relativistic perspective.  In \S 3 we investigate the perturbation theory and calculate the leading order corrections to the matter power spectrum.  We also investigate the relationship between these calculations and the consistency relations investigated in \cite{Horn:2014rta}.  In \S 4 we develop numerical simulations of the pion field evolution, and in \S 5 we measure evolution of the pion field and the EFT parameters from an N-body simulation.  Finally, we conclude in \S 6 and indicate further directions.

\section{Pion field description of LSS}

In this section we discuss the pion field formalism for LSS, originally developed by Nicolis and Vernizzi\cite{Nicolis} and further discussed in \cite{Horn:2014rta}.  For other approaches to describing the evolution of fluid perturbations in spacetime from an action principle, see e.g. \cite{Son:2002zn, Endlich:2010hf, Dubovsky:2011sj} for a general discussion of fluid actions and e.g.\cite{Rigopoulos:2014esa, Fuhrer:2015cia, Floerchinger:2016hja} for specific applications to the clustering of cold dark matter.

\subsection{Derivation from fluid equations}

One possible starting point is to consider the Navier-Stokes equations for a pressureless fluid in expanding spacetime, working in the single-stream approximation and neglecting velocity dispersion.  In terms of the conformal time $\tau$ and the comoving spatial coordinates $\vec{x}$, the background metric has the standard Friedmann-Robertson-Walker form
\begin{equation}
ds^2 = a(\tau)^2(-d\tau^2 + d\vec{x}^2)\,,
\end{equation}
where $a(\tau)$ is the scale factor as a function of conformal time.  The Friedmann equation for a $\Lambda$ + cold dark matter ($\Lambda$CDM) universe is given by
\begin{equation}
\mathcal{H}^2 = \left(\frac{a'}{a}\right)^2 = \frac{8 \pi G_N}{3}\bar{\rho}a^2 = H_{0}^{2}a^2\left(\frac{\Omega_{m0}}{a^3} + (1-\Omega_{m0})\right)\,,
\end{equation}
where $G_N$ is Newton's constant, $\bar{\rho} = \bar{\rho}_0/a^3$ is the average matter density, and $H_0$, $\bar{\rho}_0$ and $\Omega_{m0}$ are respectively the Hubble expansion parameter, the matter density and amount of matter as a percentage of the critical density in the present epoch $a = 1$.  Restricting to the matter dominated regime $\Omega_{m} = 1$, we have $a(\tau) \propto \tau^2$, and for general values of $\Omega_{m0}$, $\tau(a)$ can be solved in terms of hypergeometric functions as.  
\begin{equation}
\tau(a) = \frac{2\sqrt{a}}{H_0 \sqrt{\Omega_{m0}}}{}_{2}F_{1}\left(\frac{1}{6}, \frac{1}{2}, \frac{7}{6}, a^3(1-\Omega_{m0}^{-1})\right)\,.
\end{equation}
By taking derivatives of the Friedmann equation we can derive the following useful expressions:
\begin{equation}
\mathcal{H}' = \Big(1 - \frac{3}{2}\Omega_m\Big)\mathcal{H}^2\,, \qquad \mathcal{H}'' = \mathcal{H}\mathcal{H}' + \mathcal{H}^3\,,
\end{equation}
in the $\Lambda$CDM universe.

Perturbations in the matter density can by expressed in terms of the overdensity $\delta = (\rho - \bar{\rho})/\bar{\rho}$, the velocity field $\vec{v}$, and the Newtonian potential $\Phi$.  The perturbed metric becomes
\begin{equation}
ds^2 = a(\tau)^2(-(1 + 2\Phi)d\tau^2 + (1-2\Phi)d\vec{x}^2)\,,
\end{equation}
and the perturbations obey the continuity equation, the Euler equation, and the Poisson equation.  Respectively, these are:
\begin{equation}\label{fluidEqns}
\begin{split}
\delta' + \vec{\nabla}\cdot\left((1+\delta)\vec{v}\right) &= 0\,,\\
v_{i}' + \mathcal{H} v_{i} + (v^{j} \nabla_{j})v_{i} + \nabla^{j} \tau_{ij} &= -\nabla_{i}\Phi\,,\\
\nabla^2 \Phi = 4 \pi G_N \bar{\rho}a^2 \delta &= \frac{3}{2}\mathcal{H}^2 \Omega_m \delta \,,
\end{split}
\end{equation}
where $\tau_{ij}$ is the stress tensor arising from velocity dispersion in the matter fluid.  The evolution of the stress tensor depends on higher moments of the Boltzmann hierarchy; however, in the limit of vanishing velocity dispersion, the stress tensor can be self-consistently set to zero.  

For the case of gradient flow the dynamics in \eqref{fluidEqns} can be expressed in terms of a single degree of freedom,
\begin{equation}
\vec{v} = \vec{\nabla}\pi\,.
\end{equation}
To see that $\vec{\omega} = \vec{\nabla} \times \vec{v}$ can be self-consistently set to zero, take the curl of the Euler equation to find 
\begin{equation}\label{vorticity}
\vec{\omega}' + \mathcal{H}\vec{\omega} + \vec{\omega}(\vec{\nabla} \cdot \vec{v}) - (\vec{\omega} \cdot \vec{\nabla})\vec{v} + (\vec{v}\cdot \vec{\nabla})\cdot\vec{\omega} = 0\,.
\end{equation}
Therefore, if the vorticity vanishes initially, it will remain zero if the stress tensor vanishes.  With $\vec{v} = \vec{\nabla}\pi$, the Euler equation can be integrated to give
\begin{equation}
\Phi = - \Bigg(\pi' + \mathcal{H}\pi + \frac{1}{2}(\nabla \pi)^2 \Bigg)\,,
\end{equation}
and the Poisson equation then gives
\begin{equation}\label{pi_to_delta}
\delta = \frac{1}{\mathcal{H}' - \mathcal{H}^2}\nabla^2\Bigg(\pi' + \mathcal{H}\pi + \frac{1}{2}(\nabla \pi)^2 \Bigg)\,.
\end{equation}
Both these quantities can therefore be written as nonlinear functions of the single scalar $\pi$ field.  The continuity equation then gives us the equation of motion for $\pi$:
\begin{equation}\label{pi_eom}
\begin{split}
\pi'' + 2\mathcal{H} \pi' + &+2\mathcal{H}'\pi + \frac{1}{2}\mathcal{H}(\nabla \pi)^2 + \vec{\nabla} \pi \cdot \vec{\nabla} \pi' \\ &+  \frac{1}{\nabla^2}\vec{\nabla} \cdot \Bigg[(\vec{\nabla} \pi) \nabla^2 \Bigg(\pi' + \mathcal{H}\pi + \frac{1}{2}(\nabla \pi)^2 \Bigg)\Bigg] = 0\,.
\end{split}
\end{equation}
This single equation of motion expresses the evolution of the scalar perturbations in $\Lambda$CDM as a single degree of freedom.  The solution to the linearized equation of motion is given by
\begin{equation}
\pi_{lin}(\tau) = \frac{c_{decaying}}{a^2} + \frac{c_{growing}}{a^2}\int^{\tau} d\tau' a(\tau')^2\,.
\end{equation}
The growing mode can be expressed as a hypergeometric function (see \S 3) and is simply proportional to $\tau$ when matter dominates at early times.  

As discussed in \cite{Baumann:2010tm}, from the inflationary power spectrum, we have $\Phi \sim 10^{-5}$, and we can use the equations of motion to estimate the characteristic scale of each variable as
\begin{equation}
\Phi \sim 10^{-5}\,, \qquad v \sim \frac{k}{\mathcal{H}}10^{-5}\,, \qquad \delta \sim \frac{k^2}{\mathcal{H}^2}10^{-5}\,, \qquad \pi \sim \frac{10^{-5}}{\mathcal{H}}\,,
\end{equation} 
in the linear regime.  The nonlinear terms $\delta \vec{v}$ and $(\vec{v}\cdot \vec{\nabla}) \vec{v}$ are both suppressed on the order $k^2 \pi / \mathcal{H} \sim 10^{-5}k^2/\mathcal{H}^2$ relative to the scale of the linear terms, and the intrinsic scale of galaxy clusters is therefore set by
\begin{equation}
k_{NL}^{2} \sim 10^{5}\mathcal{H}^2\,.
\end{equation}
Given a (comoving) Hubble scale currently on the order of 4 Gpc, this implies the scale $k_{NL}^{-1} \approx$ 10 Mpc for galaxy clustering \cite{Baumann:2010tm}.  In the pion field equation of motion Eq.~\eqref{pi_eom}, nonlinear terms become important when  $k^2 \pi /\mathcal{H} \sim 1$.  In this picture it is manifest that the terms $\delta \vec{v}$ and $(\vec{v}\cdot \vec{\nabla})\vec{v}$ are of the same size.

\subsection{Derivation from relativistic action}

The pion field description can also be derived starting with the Lagrangian for a relativistic irrotational fluid \cite{Son:2002zn}, which can be derived from symmetry considerations.  Following the notation used in \cite{Horn:2014rta}, we have
\begin{equation}
S = \int d^{4}x \sqrt{-g}\Big[\frac{1}{2}M_{P}^{2}\mathcal{R} + \mathcal{P}(X)\Big]\,,
\end{equation}
where $X = \sqrt{- g^{\mu \nu}\partial_{\mu}\phi \partial_{\nu}\phi}$.  Note that the action here is also related to the fluid Lagrangian in \cite{Endlich:2010hf, Dubovsky:2011sj} via a Legendre transformation.  The stress-energy tensor
\begin{equation}
T_{\mu \nu} = 2 \mathcal{P}_{,X}\partial_{\mu}\phi\partial_{\nu}\phi + g_{\mu \nu}\mathcal{P}
\end{equation}
has the form of a perfect fluid, where the density $\rho$, the pressure $P$, and the 4-velocity $u_{\mu}$ are given by
\begin{equation}
\rho = 2 X \mathcal{P}_{,X} - \mathcal{P}\,, \qquad P = \mathcal{P}\,, \qquad u_{\mu} = \frac{-\partial_{\mu}\phi}{X}\,.
\end{equation} 
Using the ansatz $\mathcal{P}(X) \propto X^{\frac{1+w}{2w}}$
in the limit $w \to 0$, and separating out the background $\phi = \bar{\phi}- \bar{\phi}' \pi$, the 3-velocity corresponds to
\begin{equation}
\vec{v} = \frac{\partial_{i}\pi}{1-\mathcal{H}\pi - \pi'}\,,
\end{equation}
and the action to leading order in perturbations becomes
\begin{equation}
S = -\int d\tau d^{3}x\, \Bigg[\frac{1}{2}\bar{\rho}a^4 (\vec{\nabla} \pi)^2 + M_{P}^{2} \Bigg(\vec{\nabla}\Bigg[(a\pi)' + \frac{1}{2}a(\vec{\nabla}\pi)^2\Bigg]\Bigg)^2\Bigg]\,,
\end{equation}
which gives the equation of motion found in Eq.~\eqref{pi_eom}.  This description was first written down by Nicolis and Vernizzi \cite{Nicolis}, and further studied in \cite{Horn:2014rta}.  It should be noted that we have neglected corrections of higher order in $\pi' \sim \mathcal{H}\pi$, $w$, and $\Phi$, all of which are expected to be suppressed on the order of $10^{-5}$, but we are keeping corrections of higher order in $k^2/k_{NL}^{2}$.   

In the Newtonian limit, the action is invariant both under the time-dependent shift symmetry $\pi \to \pi + \epsilon$ and under the time-dependent translation symmetry
\begin{equation}\label{EFTsymms}
\begin{split}
\tau \to \tau \,, \qquad \vec{x} \to \vec{x} + \vec{n}\,, &\qquad \frac{d}{d\tau} \to \frac{d}{d\tau} - \vec{n}' \cdot \vec{\nabla}\,, \qquad \vec{\nabla} \to \vec{\nabla}\,, \\
\qquad &\pi \to \pi + \vec{n}' \cdot \vec{x}\,,
\end{split}
\end{equation}
where $\epsilon(\tau)$ and $\vec{n}(\tau)$ are further restricted to obey the adiabatic mode conditions that $\epsilon' + 2 \mathcal{H}\epsilon$ and $\vec{n}'' + 2 \mathcal{H} \vec{n}'$ be constant in time.  It was shown in \cite{Horn:2014rta} that the symmetries can be extended to diffeomorphism symmetries of the full general relativistic action, in which case the shift symmetry becomes a time-dependent time translation with a time-independent dilatation symmetry on the spatial slice, and the time-dependent translation symmetry includes a time-independent special conformal transformation of the spatial slice. The $\pi$ field can be therefore be thought of as the Goldstone boson corresponding to the breaking of spacetime symmetry by the fluid, in addition to being the Goldstone boson of the broken $U(1)$ charge of a superfluid.  As often happens in the case of spontaneously broken spacetime symmetries, the same Goldstone boson can correspond to more than one broken symmetry via inverse Higgs constraints (see e.g. \cite{Nicolis:2013sga}).

To summarize, we can write the growth of scalar perturbations in a $\Lambda$CDM universe in terms of a scalar pion field, corresponding to the spontaneous breaking of spacetime symmetry in the presence of pressureless matter.  There are several potential advantages to this picture; in particular, it simplifies the interrelated variables $\delta, \vec{v}, \Phi$ into a single scalar degree of freedom, and makes manifest that the nonlinear terms $\vec{\nabla}(\delta \vec{v})$ and $(\vec{v} \cdot \vec{\nabla})\vec{v}$ are of similar size in the effective field theory of LSS.  Furthermore, as discussed in \cite{Horn:2014rta}, consistency relations are naturally expressed in terms of the $\pi$ field.  We now analyze the EFT corrections within this framework, and show that these are very simple to derive at next to leading order.

\subsection{EFT and higher order corrections}

The symmetries in Eq.\eqref{EFTsymms} restrict additional terms in the Lagrangian to be functions of the invariant quantities
\begin{equation}
\tau \,, \qquad \nabla_{i}\nabla_{j} \pi \,, \qquad \vec{\nabla} \nabla_{i}\nabla_{j} \pi\,, \qquad \Bigg(\frac{d}{d\tau}+\vec{\nabla}\pi \cdot \vec{\nabla}\Bigg)\nabla_{i}\nabla_{j} \pi\,, \cdots
\end{equation}
plus additional terms including higher order combinations of spatial and convective time derivatives.  To leading order in the derivative expansion, there are two possible EFT terms in the Lagrangian:
\begin{equation}
\delta S = M_P^2\int d\tau d^{3}x\, a^2\Bigg[ \alpha(\tau) (\nabla^2 \pi)^2 - \beta(\tau) (\nabla^2 \pi')^2 \Bigg]\,.
\end{equation}
We have used integration by parts to eliminate terms of the form $(\nabla^2 \pi)(\nabla^2 \pi')$, and we can use the linearized equations of motion to rewrite $\pi''$ in terms of $\pi, \pi'$.  These contribute the following terms in the equation of motion Eq.~\eqref{pi_eom}:
\begin{equation}\label{pi_eom_corrected}
\begin{split}
\pi'' + 2\mathcal{H} \pi' + 2\mathcal{H}'\pi &+ \frac{1}{2}\mathcal{H}(\nabla \pi)^2 + \vec{\nabla} \pi \cdot \vec{\nabla} \pi' \\ +  \frac{1}{\nabla^2}\nabla_i \left(\nabla_i \pi \nabla^2 \Bigg[\pi' + \mathcal{H}\pi \right.&+ \left.\frac{1}{2}(\nabla \pi)^2 \Bigg]\right) - \nabla^2 \left( \tilde{c}_{s}^{2}(\tau)\pi + \frac{\tilde{c}_{v}(\tau)^{2}}{\mathcal{H}}\pi'\right)= 0\,,
\end{split}
\end{equation}
where $\tilde{c}_{s}^{2} = \alpha - 2 \mathcal{H}'\beta$ and $\tilde{c}_{v}^{2} = \mathcal{H}\beta'$.  In the parameterization of \cite{Baumann:2010tm, Carrasco:2012cv}, this is equivalent to adding the terms
\begin{equation}
\nabla^{j}\tau_{ji} = \nabla_{i}\left(c_{s}^{2}\delta - \frac{c_{v}^{2}}{\mathcal{H}} \nabla^2 \pi\right) = -\nabla_i \Bigg(\frac{c_s^2}{\mathcal{H}^2 - \mathcal{H}'}\nabla^2 \pi' + \Bigg(\frac{c_s^2\mathcal{H}}{\mathcal{H}^2 - \mathcal{H}'}+\frac{c_v^2}{\mathcal{H}}\Bigg)\nabla^2 \pi \Bigg)\,,
\end{equation}
where the untilded variables $c_{s}^{2}, c_{vis}^{2}$ are the time-dependent sound speed and (dimensionless) viscosity parameters, to the left hand side of the Euler equation.  We will use the untilded variables for the EFT coefficients in the conventional fluid picture, following the notation of \cite{Baumann:2010tm, Carrasco:2012cv}, and the tilded versions for the pion field picture.  
The relationship between the pion field EFT parameters and the fluid parameters $c_{s}^{2}, c_{vis}^{2}$ are:
\begin{equation}\label{tilded_EFT_coeffs}
\begin{split}
\tilde{c}_{s}^{2} = (\alpha - 2 \mathcal{H}' \beta) &= \Bigg(\frac{{c_{s}^{2}}'\mathcal{H}+{c_{s}^{2}}(2\mathcal{H}^2-{\mathcal{H}}')}{{\mathcal{H}}^{2}-\mathcal{H}'} + \frac{{c_{v}^{2}}'\mathcal{H}+{{c_{v}^{2}}}(\mathcal{H}^2-{\mathcal{H}}')}{\mathcal{H}^2}\Bigg)\,,\\
\frac{\tilde{c}_{v}^{2}}{\mathcal{H}} &= \beta' = \Bigg(\frac{{c_{s}^{2}}'+\mathcal{H}{c_{s}^{2}}}{{\mathcal{H}}^{2}-\mathcal{H}'} + \frac{{c_{v}^{2}}}{\mathcal{H}}\Bigg) \,.
\end{split}
\end{equation}
The time dependence of the EFT coefficients is not fixed by the top-down symmetries alone; nevertheless, we can estimate it by considering that the physical origin of the sound speed and viscosity is the growth of velocity dispersion as structures collapse and virialize on small scales.  In this case we can estimate that the dimensionless sound speed is
\begin{equation}
c_{s}^{2} \sim v_{p}^{2}/c^2\,,
\end{equation}
i.e., the physical sound speed is of the same scale as the peculiar velocity at the time of collapse, and
\begin{equation}
c_{v}^{2} \sim \frac{\mathcal{H}}{\bar{\rho}_m c^2}\left(\frac{4}{3}\eta + \zeta \right) \sim \frac{\mathcal{H}\bar{\rho}_m v_{p}^{2} \tau}{\bar{\rho}_m c^2} \sim \mathcal{H} \tau v_{p}^2 \propto \frac{1}{\sqrt{\Omega_m}} \frac{v_{p}^{2}}{c^{2}}
\end{equation}
for the dimensionless viscosity parameter.  Here $\eta$ and $\zeta$ are the shear and bulk viscosity in conventional units, and in the last step we have used the approximation $\mathcal{H} \tau \propto \Omega_m^{-1/2}$ in a $\Lambda$CDM universe.  Since the peculiar velocities are generated by the collapse of structures, we estimate $v_p \propto D$ at the time of collapse, where $D(\tau)$ is the linear growth factor of the $\pi$ field.  This gives $c_{s}^{2} \propto D^2$, $c_{v}^{2} \propto \Omega_{m}^{-1/2} D^2$, which can be shown to give a good approximation to the time dependence found from perturbation theory in \cite{Carrasco:2012cv}.


So far the EFT terms to NLO are very simple and give local corrections to the non-local equation of motion in Eq.\eqref{pi_eom}.  The NNLO corrections in the derivative expansion are more complicated, and include terms in the Lagrangian such as
\begin{equation}
\begin{split}
&\nabla^2 \pi \nabla^2(\nabla^2 \pi)\,, \qquad \nabla_i \nabla_j \pi \nabla^2(\nabla_i \nabla_j \pi)\,, \\
(\nabla^2 \pi)^3\,, \qquad &(\nabla_{i}\nabla_{j}\pi)^2 \nabla^2 \pi \,, \qquad (\nabla_{i}\nabla_{j}\pi)(\nabla_{j}\nabla_{k}\pi)(\nabla_{k}\nabla_{i}\pi)\,.
\end{split}
\end{equation}
These may come together with time-dependent coefficients, and one or more of the $\pi$ fields may be replaced by $\pi'$.  Accounting for integration by parts in time and space, there are a total of 17 independent terms.  There also also four more independent terms arising from the convective part of the time derivatives, namely
\begin{equation}
\begin{split}
(\nabla_{i}\nabla_{j}\pi)(\vec{\nabla}\pi \cdot \vec{\nabla})( \nabla_{i}\nabla_{j}\pi')\,, \qquad (\nabla_{i}\nabla_{j}\pi)(\vec{\nabla}\pi' \cdot \vec{\nabla})( \nabla_{i}\nabla_{j}\pi')\,,\\
(\nabla^{2}\pi)(\vec{\nabla}\pi \cdot \vec{\nabla})(\nabla^{2}\pi')\,, \qquad (\nabla^{2}\pi)(\vec{\nabla}\pi' \cdot \vec{\nabla})(\nabla^{2}\pi')\,.
\end{split}
\end{equation}
These terms will give rise to a variety of possible NNLO corrections to Eq.~\eqref{pi_eom_corrected}, most of which are nonlocal once the inverse Laplacian operator is applied.  Nevertheless, if the Lagrangian is local, the inverse Laplacian operator will be the only source of nonlocality in the equations of motion.


\section{Perturbation theory and the power spectrum}

We can expand the equation of motion Eq.~\eqref{pi_eom_corrected} in the derivative expansion, and solve perturbatively for the $\pi$ field as $\pi = \pi_1 + \pi_2 + \pi_3 + \cdots$.  The solution to the linearized equation of motion $\pi_1'' + 2\mathcal{H}\pi_1' + 2\mathcal{H}'\pi_1 = 0$ is given by
\begin{equation}
\pi_{1}(\tau) =  \frac{c_{1}}{a^2}\int^{\tau} d\tau' a(\tau')^2 + \frac{c_{2}}{a^2}\,.
\end{equation}
The time dependence of the growing mode can be solved in terms of the scale factor as
\begin{equation}
\begin{split}
D(a) &= \frac{5}{a^2}\int^{\tau} a(\tau')^2 d\tau' = \frac{5}{a^2}\int^{a}\frac{da'}{H_0 \sqrt{\frac{\Omega_{m0}}{a'^3}+(1-\Omega_{m0})}} \\
&= \frac{2\sqrt{a}}{H_0 \sqrt{\Omega_{m0}}}{}_{2}F_{1}\Big(\frac{1}{2},\frac{5}{6}, \frac{11}{6}, a^3(1-\Omega_{m0}^{-1})\Big)\,.
\end{split}
\end{equation}
Here we have chosen the overall normalization so that for early times $a \ll 1$, this expression reduces to
\begin{equation}
D(a \ll 1) \approx \frac{2\sqrt{a}}{H_0 \sqrt{\Omega_{m0}}} \approx \tau\,.
\end{equation}
The hypergeometric function will reach a local maximum around the time of matter-$\Lambda$ equality.  For $\Omega_{m0} = 0.31$, the exact value of the local maximum is $2.67/H_0$ and it is reached around $a = 0.872$.

The Green's function solving $\pi'' + 2\mathcal{H}\pi' + 2\mathcal{H}'\pi = \delta(\tau-\tau')$ is given by
\begin{equation}\label{Green_func}
G(\tau, \tau') = \left(\frac{D(a(\tau))}{5} - \frac{a(\tau')^2}{a(\tau)^2}\frac{D(a(\tau'))}{5}\right)\theta(a-a')\,,
\end{equation}
where $\theta$ is the Heaviside theta function.  In the matter-dominated regime the Green's function simplifies to
\begin{equation}
G(\tau, \tau_0) = \Big(\frac{\tau}{5} - \frac{\tau_{0}^{5}}{5 \tau^4}\Big)\theta(\tau-\tau_0)\,.
\end{equation}

To leading order the power spectrum can be estimated parametrically by
\begin{equation}
\begin{split}
\langle \tilde{\pi}(\vec{k})\tilde{\pi}(\vec{k}')\rangle &= P_{0}(k)(2\pi)^{3}\delta^{3}(\vec{k}+\vec{k}') \\
&\sim (2 \pi)^3 \frac{10^{-10}}{\mathcal{H}^2 k^3} (2\pi)^{3}\delta^{3}(\vec{k}+\vec{k}') \qquad k \lesssim k_{mr}\\
&\sim (2 \pi)^3 \frac{10^{-10}}{\mathcal{H}^2 k^3}\Bigg(\frac{k_{mr}}{k}\Bigg)^4 \Big(\log(k/k_{mr})\Big)^2(2\pi)^{3}\delta^{3}(\vec{k}+\vec{k}') \qquad k \gtrsim k_{mr}
\end{split}
\end{equation}
Here $k_{mr} \sim \mathcal{H}_{mr} = \sqrt{2\Omega_m^{2}/\Omega_r}(H_0/c) \approx 2\pi/(600 Mpc)$ corresponds to the scale of modes that reenter the horizon around the era of matter-radiation equality.  On smaller scales, the factor $(k/k_{mr})^4 (\log(k/k_{mr}))^2$ arises due to the lag time between horizon reentry and matter-radiation equality, at which time the modes start growing.  For modes that reenter the horizon in the radiation dominated regime, the nonlinear corrections will depend on both $k_{NL} \sim 10^{5/2}\mathcal{H}$ and on $k_{mr}$, and it is therefore of interest to perform a more careful perturbative calculation of the power spectrum to NLO.

\subsection{NLO power spectrum in matter-dominated regime}

Starting with the growing linearized mode $\pi_{1}(\vec{x}, \tau) = \hat{\pi}_{1}(x) D(\tau)$, we can solve the $\pi$ equation of motion perturbatively as
\begin{equation}
\pi = \pi_1 + \pi_2 + \pi_3 + \cdots
\end{equation}
where
\begin{equation}
\begin{split}
\pi_{2}'' + 2\mathcal{H}\pi_2' + 2\mathcal{H}'\pi_2 &= -\frac{1}{2}\mathcal{H} (\vec{\nabla}\pi_{1})^2 - (\vec{\nabla}\pi_{1}\cdot \vec{\nabla} \pi_{1}') \\ &- \frac{1}{\nabla^2}\vec{\nabla}\cdot\left(\vec{\nabla}\pi_{1} \nabla^2 \left[\pi_{1}' + \mathcal{H}\pi_{1}\right]\right) + \tilde{c}_{s}^{2}\nabla^2 \pi_1 + \frac{\tilde{c}_{v}^{2}}{\mathcal{H}}\nabla^2 \pi'_1 \,,\\
\pi_{3}'' + 2\mathcal{H}\pi_3' + 2\mathcal{H}'\pi_3 &= -\mathcal{H} (\vec{\nabla}\pi_{1}\cdot \vec{\nabla}\pi_2) - (\vec{\nabla}\pi_{1}\cdot \vec{\nabla} \pi_{2}')-(\vec{\nabla}\pi_{2}\cdot \vec{\nabla} \pi_{1}') \\ - \frac{1}{\nabla^2}\vec{\nabla}\cdot \Bigg(\vec{\nabla}\pi_{2} &\nabla^2 \left[\pi_{1}' + \mathcal{H}\pi_{1}\right] +\vec{\nabla}\pi_{1} \nabla^2 \left[\pi_{2}' + \mathcal{H}\pi_{2} + \frac{1}{2}(\nabla \pi_1)^2\right]\Bigg) + \cdots \,,\\
\end{split}
\end{equation}

Restricting ourselves initially to the matter-dominated regime $\Omega_m = 1$ and $\tilde{c}_{s}^{2} = \tilde{c}^{2}_{v} = 0$, we can factor out the (conformal) time dependence to write the $\pi$ field perturbatively as
\begin{equation}
\pi = \hat{\pi}_1 (\vec{x}) \tau + \hat{\pi}_2 (\vec{x}) \tau^3 + \hat{\pi}_3 (\vec{x}) \tau^5 + \cdots\,,
\end{equation}
where
\begin{equation}
\hat{\pi}_2 = -\frac{1}{7}(\vec{\nabla} \hat{\pi}_1)^2 - \frac{3}{14}\frac{1}{\nabla^2}(\vec{\nabla}\hat{\pi}_1 \cdot \nabla^2 \hat{\pi}_1) + \cdots
\end{equation}
contains terms coming from the three-way vertex diagram, and the next term 
\begin{equation}
\hat{\pi}_3 = - \frac{1}{6}\vec{\nabla}\hat{\pi}_1 \cdot \vec{\nabla}\hat{\pi}_2 - \frac{1}{36}\frac{1}{\nabla^2}\vec{\nabla}\left(3 \vec{\nabla}\hat{\pi}_2 \nabla^2 \hat{\pi}_1 + 5 \vec{\nabla} \tilde{\pi}_1\nabla^2 \hat{\pi}_2 + \frac{1}{2}\vec{\nabla}\hat{\pi}_1 (\vec{\nabla} \hat{\pi}_1)^2\right)
\end{equation}
includes terms from the three-way as well as the four-way vertex.  For a given cosmic time, the nonlocal inverse Laplacian operator needs to be evaluated only once for each perturbation step.

\begin{figure}[htbp]
\begin{center}
\includegraphics[scale=0.6]{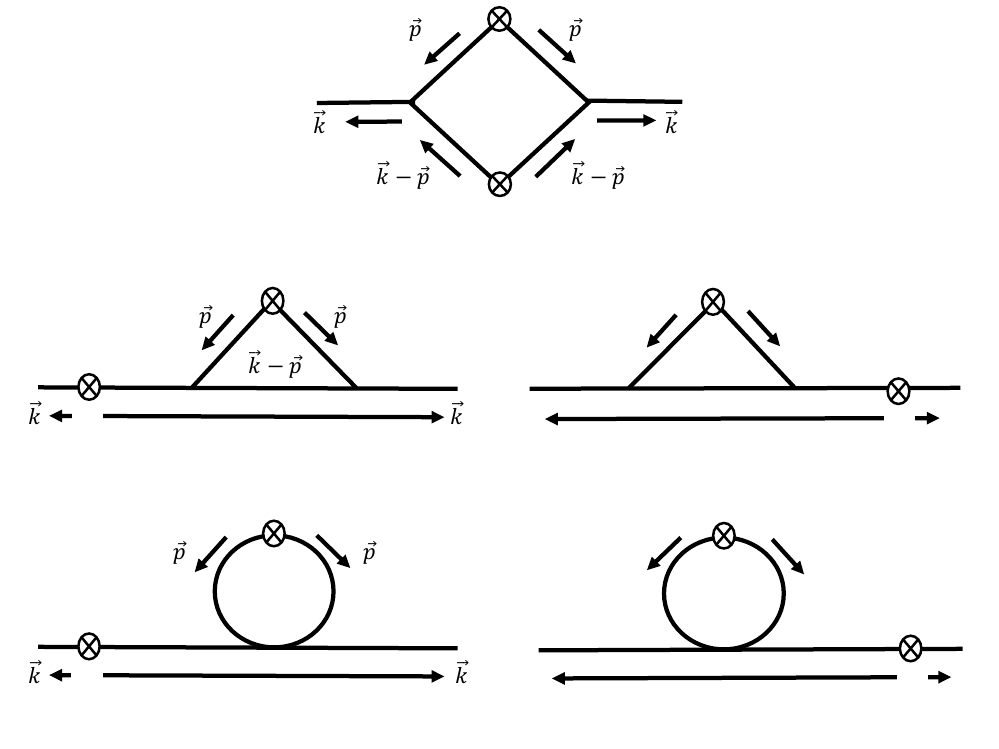}
\caption{The diagrammatic expansion for the power spectrum at next-to leading order.  The arrows indicate the direction of time evolution and the circles with crosses indicate the linearized power spectrum for the initial conditions.}\label{NLO_Pk_diagrams}
\end{center}
\end{figure}

The next to leading order terms in the power spectrum come from the terms
\begin{equation}
\langle \tilde{\pi}_2(\vec{k}, \tau) \tilde{\pi}_2 (\vec{k}', \tau) \rangle + \langle \tilde{\pi}_3(\vec{k}, \tau) \tilde{\pi}_1 (\vec{k}', \tau) \rangle + \langle \tilde{\pi}_1(\vec{k}, \tau) \tilde{\pi}_3 (\vec{k}', \tau) \rangle
\end{equation}
which include the diagrammatic contributions found in Fig.~\ref{NLO_Pk_diagrams}.

\noindent Considering the first of these diagrams, we have a contribution
\begin{equation}
\begin{split}
\int \frac{d^{3}\vec{p}}{(2\pi)^3}&\frac{1}{2}\left(\frac{2}{7}\vec{p}\cdot(\vec{k}-\vec{p}) + \frac{3}{14}(\vec{p}\cdot\vec{k})\frac{(k-p)^2}{k^2} + \frac{3}{14}(k^2 - \vec{p}\cdot\vec{k})\frac{p^2}{k^2}\right)^2 \\
&\times \hat{P}_{0}(|k-p|)\hat{P}_{0}(p)\tau^6\,,
\end{split}
\end{equation}
where $P_0(k, \tau) = \hat{P}_0(k)\tau^2$ is the power spectrum corresponding to the growing linearized modes.  The second pair of diagrams each give rise to a contribution
\begin{equation}
\begin{split}
2\int \frac{d^{3}\vec{p}}{(2\pi)^3}&\left(\frac{1}{6}\vec{p}\cdot(\vec{k}-\vec{p}) + \frac{5}{36}(\vec{p}\cdot\vec{k})\frac{(k-p)^2}{k^2} + \frac{1}{12}(k^2 - \vec{p}\cdot\vec{k})\frac{p^2}{k^2}\right)\\
&\times \left(-\frac{2}{7}(\vec{p}\cdot\vec{k}) + \frac{3}{14}\frac{\vec{k}\cdot(\vec{k}-\vec{p})p^2 + \vec{p}\cdot(\vec{p}-\vec{k})k^2}{(k-p)^2}\right)\hat{P}_{0}(p)\hat{P}_{0}(k)\,,
\end{split}
\end{equation}
and the third pair contributes
\begin{equation}
\begin{split}
2\int \frac{d^{3}\vec{p}}{(2\pi)^3}&\left(\frac{1}{36}(\vec{p}\cdot\vec{k})^2 \frac{(k-p)^2}{k^2}\right)\hat{P}_{0}(p)\hat{P}_{0}(k)\,.
\end{split}
\end{equation}

To solve the numerical integral, we can approximate the power spectrum $P_{0}(k,\tau) = \hat{P}_{0}(k)\tau^2$ by making use of the fitting function in \cite{Eisenstein:1997ik} for the matter power spectrum, together with the normalization given by \cite{Bunn:1996da}.  Plotting this ansatz for $P_{0}(k, \tau)$ together with simulated output linear matter power spectrum from CLASS \cite{Blas:2011rf}, with the overall normalization constant rescaled so as to match in the $k \to 0$ limit, the two curves agree to within about 10 per cent up to $k \lesssim 100 h/Mpc$.  

\begin{figure}[htbp]
\begin{center}
\includegraphics[scale=0.6]{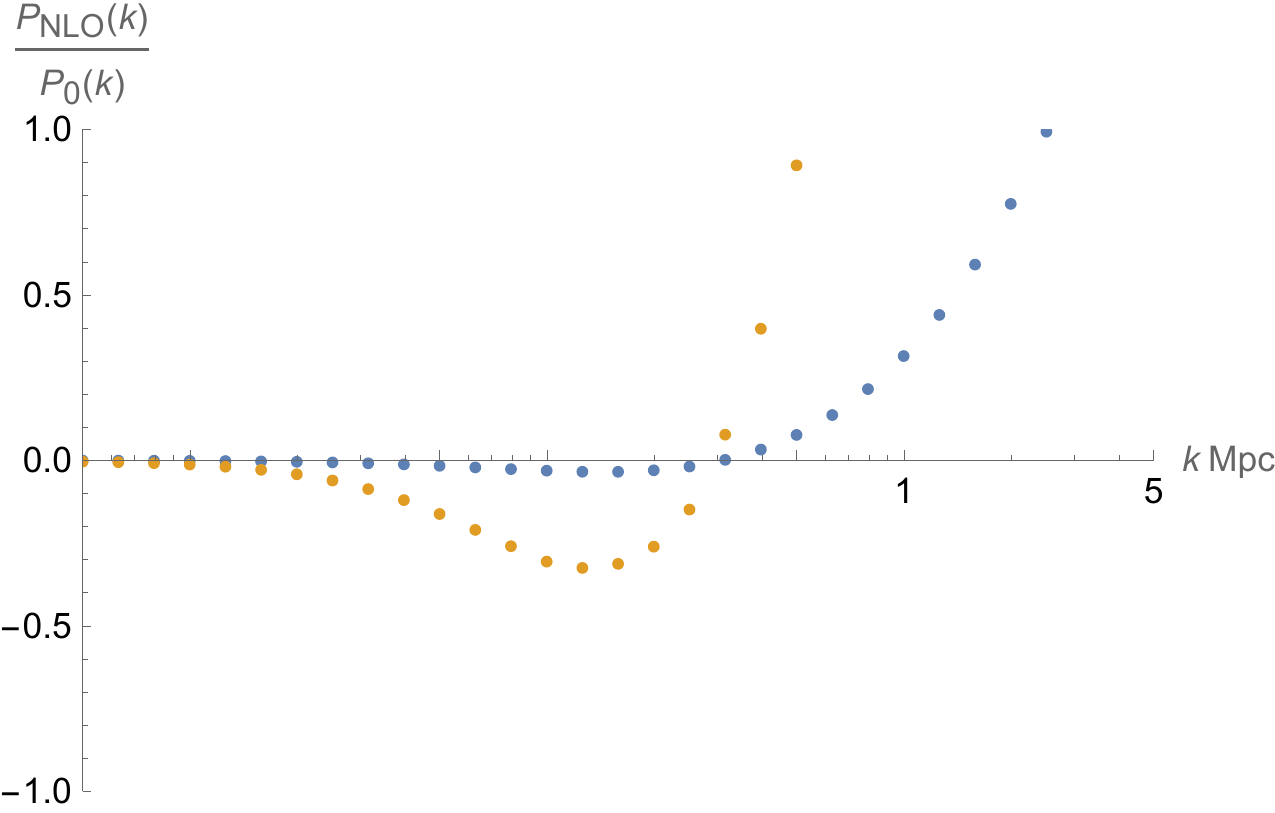}
\caption{The relative sizes of the perturbative NLO corrections to the $\pi$ field power spectrum, for the $\Lambda$CDM universe with $\Omega_{m0} = 0.308$, $H_0 = 68$kms/s/Mpc.  Blue = redshift $z = 3$, Orange = redshift $z = 0$.}
\label{NLO_corrections}
\end{center}
\end{figure}

The numerical solution for the size of the fractional next-to-leading order corrections $\Delta P_{NLO}(k,\tau)/P_{0}(k,\tau)$ to the power spectrum is plotted for $a = 0.25$, within the matter-dominated regime for the benchmark model, in Fig.~\ref{NLO_corrections}.
The scale of the graph is proportional to $\tau^4 $ and the shape of the graph depends only on $k$ and $k_{mr}$.  For our benchmark model with $\Omega_{m0} = 0.31, \Omega_r = 9.0 \times 10^{-5}$, the local minimum and the zero intercept are observed to occur around $k \sim 15 k_{mr}, 30 k_{mr}$, indicating that for a given $k$, the dominant contributions to the integrals occur for $p$ about an order of magnitude below $k$.  

For $k \gtrsim k_{mr}$, the corrections are parametrically of the scale
\begin{equation}
\begin{split}
\frac{\Delta P_{NLO}(k, \tau)}{P_{0}(k, \tau)} &\sim k^3 k^4 (2\pi)^3 \frac{10^{-10}}{\mathcal{H}^2 k^3}\left(\frac{k_{mr}}{k}\right)^{4}\log(k/k_{mr})^{2}\frac{1}{\mathcal{H}^2} \\
&\sim (2\pi)^3\frac{k_{mr}^4}{k_{NL}^{4}}(\log(k/k_{mr}))^{2}\,.
\end{split}
\end{equation}
Empirically, we find that this estimate works well for high redshift and large $k$.  For $k \lesssim k_{mr}$, on the other hand, we have
\begin{equation}
\begin{split}
\frac{\Delta P_{NLO}(k, \tau)}{P_{0}(k, \tau)} &\sim k^3 k^4 (2\pi)^3 \frac{10^{-10}}{\mathcal{H}^2 k^3}\frac{1}{\mathcal{H}^2} \\
&\sim (2\pi)^3\frac{k^4}{k_{NL}^{4}}\,.
\end{split}
\end{equation}
This is consistent with the argument given in \cite{Mercolli:2013bsa} (see also \cite{Zeldovich:1965gev, Peebles}) that the corrections to the power spectrum start at $O(k/k_{NL})^4$ in the $k \to 0$ limit, due to mass and momentum conservation.

\begin{figure}[htbp]
\begin{center}
\includegraphics[scale=0.6]{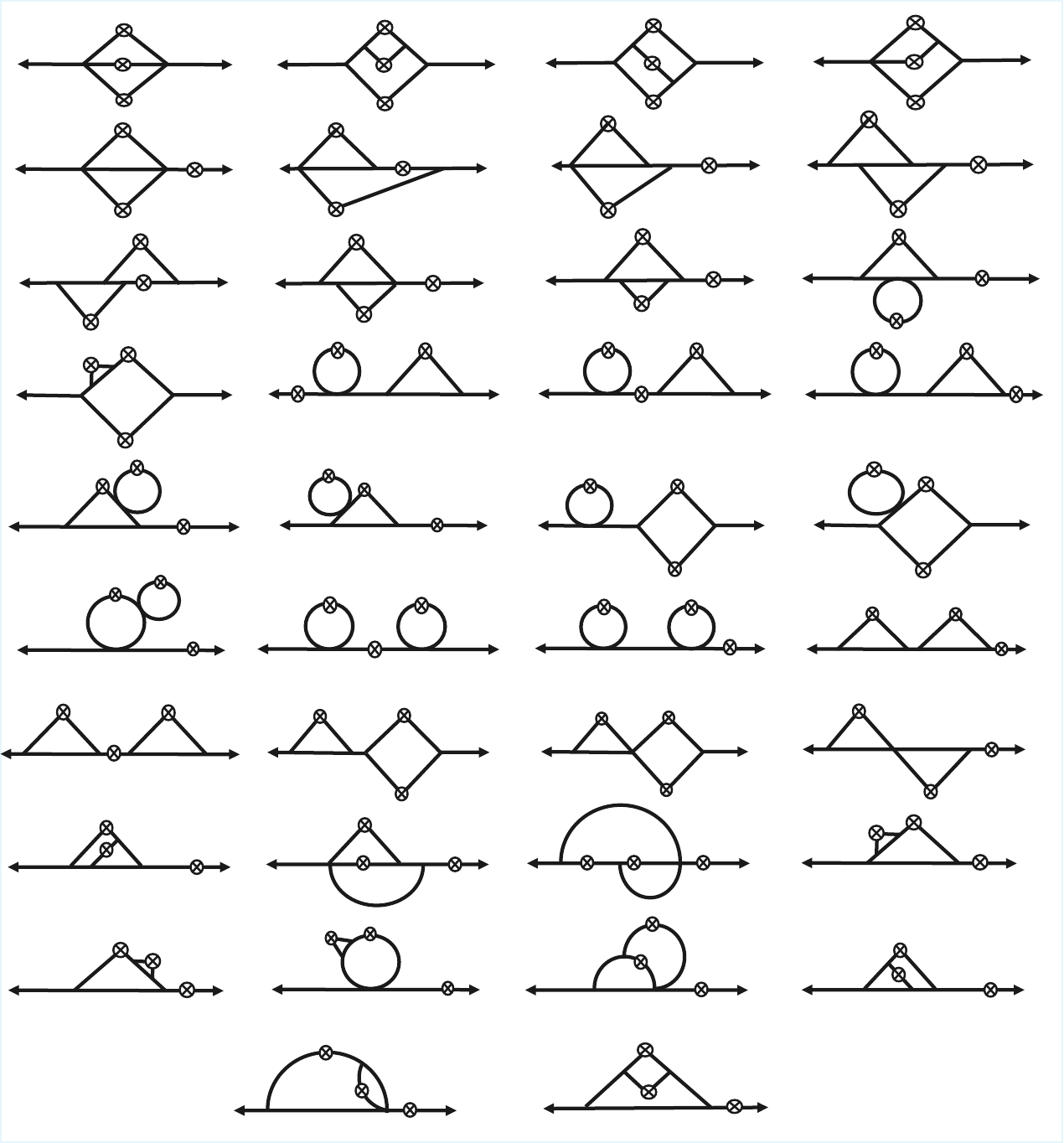}
\caption{Two-loop diagrammatic contributions to the pion field power spectrum.}
\label{NNLO_corrections}
\end{center}
\end{figure}

We can in principle continue this calculation to next-to-next-to-leading order (NNLO), using the diagrams in Fig.~\ref{NNLO_corrections}.
We can estimate the parametric scale of the NNLO contributions using $p \sim k$ within the corresponding integrals to estimate the leading contribution.  For $k \lesssim k_{mr}$, we have
\begin{equation}
\begin{split}
\frac{\Delta P_{NNLO}(k, \tau)}{P_{0}(k, \tau)} &\sim \left(k^3 (2\pi)^3 \frac{10^{-10}}{\mathcal{H}^2 k^3}\right)^{2}\frac{k^8}{\mathcal{H}^4} \\
&\sim (2\pi)^6\frac{k^{8}}{k_{NL}^{8}}\,,
\end{split}
\end{equation}
and for $k \gtrsim k_{mr}$, the functional form of the power spectrum is given by
\begin{equation}
\begin{split}
\frac{\Delta P_{NNLO}(k, \tau)}{P_{0}(k, \tau)} &\sim \left(k^3 (2\pi)^3 \frac{10^{-10}}{\mathcal{H}^2 k^3}\left(\frac{k_{mr}}{k}\right)^4\log(k/k_{mr})^{2}\right)^{2}\frac{k^8}{\mathcal{H}^4} \\
&\sim (2\pi)^6\frac{k_{mr}^{8}}{k_{NL}^{8}}\Big(\log(k/k_{mr})\Big)^{4}\,.
\end{split}
\end{equation}



\subsection{NLO corrections in $\Lambda$CDM}

Continuing to redshifts after matter-$\Lambda$ equality, the time dependence will be altered by the onset of dark energy domination.  

Writing $\pi_{1}(\vec{x}, \tau) = \hat{\pi}_{1}(\vec{x})D(\tau)$, the general solution for $\pi_2$ is given by
\begin{equation}\noindent
\begin{split}
\pi_2 (\vec{x}, \tau)
& = -\int^{\tau} d\tau' \, G(\tau, \tau')\Bigg[\left(\frac{1}{2}\mathcal{H}D^2 + D D'\right)(\vec{\nabla}\hat{\pi}_1)^2 \\ &+\left(\mathcal{H}D^2 + D D'\right)\frac{1}{\nabla^2}\vec{\nabla} \cdot (\vec{\nabla} \hat{\pi}_1 \nabla^2 \hat{\pi}_1)\Bigg]\\
&\equiv -D_{2a}(\tau) \frac{1}{2}(\vec{\nabla}\hat{\pi}_1)^2 - D_{2b}(\tau)\frac{1}{\nabla^2}\vec{\nabla} \cdot (\vec{\nabla} \hat{\pi}_1 \nabla^2 \hat{\pi}_1)\\
&\equiv D_{2a}(\tau)\hat{\pi}_{2a}(\vec{x}) + D_{2b}(\tau)\hat{\pi}_{2b}(\vec{x})\,,
\end{split}
\end{equation}
and the solution for $\pi_3$ is given by
\begin{equation}
\begin{split}
\pi_3 (\vec{x}, \tau) &= - \int^{\tau} d\tau'\, G(\tau, \tau') \Bigg[\left(\mathcal{H}D D_{2} + D'D_{2} + D D_{2}'\right)\vec{\nabla}\hat{\pi}_1 \cdot \vec{\nabla}\hat{\pi}_2 \\
&+ \left(\mathcal{H}D D_2 + D D_2'\right)\frac{1}{\nabla^2}\vec{\nabla}\left( \vec{\nabla}\hat{\pi}_1 \nabla^2 \hat{\pi}_2\right) \\
&+ \left(\mathcal{H}D D_2 + D' D_2\right) \frac{1}{\nabla^2}\left(\vec{\nabla} \tilde{\pi}_2\nabla^2 \hat{\pi}_1\right) + D^3 \frac{1}{\nabla^2}\vec{\nabla}\cdot\left(\frac{1}{2}\vec{\nabla}\hat{\pi}_1 \left(\vec{\nabla} \hat{\pi}_1\right)^2\right)\Bigg]\\
&\equiv -D_{3a}(\tau)\vec{\nabla}\hat{\pi}_1 \cdot \vec{\nabla}\hat{\pi}_2 - D_{3b}(\tau)\frac{1}{\nabla^2}\vec{\nabla}\left( \vec{\nabla}\hat{\pi}_1 \nabla^2 \hat{\pi}_2\right) \\
&\qquad - D_{3c}(\tau)\frac{1}{\nabla^2}\vec{\nabla}\left( \vec{\nabla}\hat{\pi}_2 \nabla^2 \hat{\pi}_1\right) - D_{4}(\tau)\frac{1}{\nabla^2}\vec{\nabla}\cdot\left(\frac{1}{2}\vec{\nabla}\hat{\pi}_1 \left(\vec{\nabla} \hat{\pi}_1\right)^2\right)\,,
\end{split}
\end{equation}
where the product $D_2(\tau)\hat{\pi}_{2}(\vec{x})$ implies a sum $D_{2a}\hat{\pi}_{2a} + D_{2b}\hat{\pi}_{2b}$.  The perturbative calculation expressed in Fig.~\ref{NLO_Pk_diagrams} may be performed as before, but now with time-dependent numerical coefficients.  For $\Omega_{m0} = 0.308$, $a = 1$, the square one-loop diagram gives a contribution to the $\pi$ field power spectrum which is equal to
\begin{equation}\label{PT_LCDM_one}
\begin{split}
\int \frac{d^{3}\vec{p}}{(2\pi)^3}\frac{1}{2}&\left(0.4074\vec{p}\cdot(\vec{k}-\vec{p}) + 0.3179(\vec{p}\cdot\vec{k})\frac{(k-p)^2}{k^2} + 0.3179(k^2 - \vec{p}\cdot\vec{k})\frac{p^2}{k^2}\right)^2 \\
&\times \hat{P}_{0}(|k-p|)\hat{P}_{0}(p)D^6\,,
\end{split}
\end{equation}
the two triangle diagrams give
\begin{equation}\label{PT_LCDM_two}
\begin{split}
&\qquad \qquad 2\int \frac{d^{3}\vec{p}}{(2\pi)^3}\Bigg[\left(0.0975\vec{p}\cdot(\vec{k}-\vec{p}) + 0.0851(\vec{p}\cdot\vec{k})\frac{(k-p)^2}{k^2}\right.\\
+ &\left. 0.0518(k^2 - \vec{p}\cdot\vec{k})\frac{p^2}{k^2}\right)\left(-\vec{p}\cdot\vec{k}\right)+\left(0.0745\vec{p}\cdot(\vec{k}-\vec{p}) + 0.0651(\vec{p}\cdot\vec{k})\frac{(k-p)^2}{k^2}\right.\\
&\qquad +\left. 0.0392(k^2 - \vec{p}\cdot\vec{k})\frac{p^2}{k^2}\right)\left(\frac{\vec{k}\cdot(\vec{k}-\vec{p})p^2 + \vec{p}\cdot(\vec{p}-\vec{k})k^2}{(k-p)^2}\right)\Bigg]\hat{P}_{0}(p)\hat{P}_{0}(k)D^6\,,
\end{split}
\end{equation}
and the circle diagrams give
\begin{equation}\label{PT_LCDM_three}
\begin{split}
\int \frac{d^{3}\vec{p}}{(2\pi)^3}&\left((0.1182)(\vec{p}\cdot\vec{k})^2 \frac{(k-p)^2}{k^2}\right)\hat{P}_{0}(p)\hat{P}_{0}(k)D^6\\
\end{split}\,.
\end{equation}
Numerically calculating the fractional corrections $\Delta P_{NLO}(k, \tau)/P_{0}(k,\tau)$ to the power spectrum gives a qualitatively very similar shape as in the matter-dominated epoch, plotted in Fig.~\ref{NLO_corrections}.

\subsection{EFT term corrections}

At small scales and late times, the single-stream approximation breaks down as structures collapse and virialize, and both vorticity and velocity dispersion are generated.  Nevertheless, as discussed in the previous section, the effect of short-scale dynamics on the evolution of large scales can be parameterized by the EFT coefficients.  The precise value for these coefficients will depend on the coarse graining scale and the cutoff procedure used (see e.g. \cite{Carrasco:2012cv, Pajer:2013jj}).  

The EFT coefficients contribute the additional diagrams depicted in Fig.~\ref{Pk_EFT_diagrams} to the two-point function, and these contribute to $\pi_{2}(\vec{x}, \tau)$ as:
\begin{figure}[htbp]
\begin{center}
\includegraphics[scale=0.8]{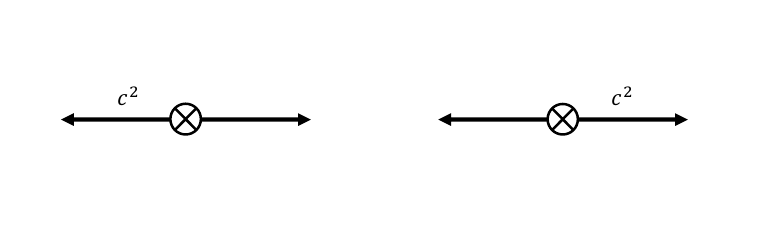}
\caption{Feynman diagrams from the inclusion of $\tilde{c}_{comb}^{2}$ terms.  The label $\tilde{c}^2$ indicates the contribution to the field evolution sourced by the EFT terms.}
\label{Pk_EFT_diagrams}
\end{center}
\end{figure}
\begin{equation}
\Delta \pi_2 = \int^{\tau} d\tau'\, G(\tau, \tau') \tilde{c}_{s}^{2}(\tau') \nabla^2 \pi_1 + \int^{\tau} d\tau'\, G(\tau, \tau') \frac{\tilde{c}_{v}^{2}(\tau')}{\mathcal{H}(\tau')} \nabla^2 \pi'_1\,.
\end{equation}
The corresponding contributions to the power spectrum are therefore
\begin{equation}
\begin{split}
\Delta P_{NLO}(k)/P_{LO}(k) &= -2k^2 \int^{\tau} d\tau'\, G(\tau, \tau') \left(\tilde{c}_{s}^{2}(\tau')D(\tau') + \frac{\tilde{c}_{v}^{2}(\tau')}{\mathcal{H}}D'(\tau')\right) /D(\tau)\\
&= -2k^2 \int^{\tau} d\tau'\, G(\tau, \tau') \tilde{c}_{comb}^{2}(\tau')D(\tau')/D(\tau)\,,\\
\end{split}
\end{equation}
where $\tilde{c}_{comb}^{2} = \tilde{c}_{s}^{2} + (D'/D)\tilde{c}_{v}^{2}$ is the combination of EFT coefficients appearing in the power spectrum.  Note that a positive value for this coefficient tends to dampen the growth of the power spectrum.

The time dependence of the EFT coefficients can be estimated using the back-of-the-envelope calculations in \S 2.3 as $\tilde{c}_{comb}^{2} \propto D^2$.  
Using Eq.~\eqref{tilded_EFT_coeffs} and investigating the time dependencies numerically for the benchmark model, $\tilde{c}^{2}_{comb}(\tau)$ is proportional to $D(\tau)^2$ within a 10 percent discrepancy before the epoch of matter-$\Lambda$ equality, and within a 20 percent discrepancy between matter-$\Lambda$ equality and the present day.  Since most of structure formation on large scales takes place at redshifts $z \gtrsim 1$, we expect the approximation $\tilde{c}_{comb}^{2} \propto D^2$ will give a reasonably precise calculation of the correction to the power spectrum.  Using this ansatz, we have
\begin{equation}\label{cs2_corrections_2pt}
\begin{split}
\Delta P_{NLO}(k)/P_{0}(k) &= -2k^2 \int^{\tau} d\tau'\, G(\tau, \tau') \tilde{c}_{comb}^{2}(\tau')D(\tau')/D(\tau)\\
&\approx -\left(\frac{k}{0.36\textrm{Mpc}^{-1}}\right)^2 \left(\frac{\tilde{c}_{comb,0}^{2}}{10^{-6}}\right) \left(\frac{D(\tau)}{D(\tau_0)}\right)^{4}\,.
\end{split}
\end{equation}


Putting the field theory corrections all together, we have contributions from both gravitational collapse and from the EFT coefficients, which are respectively suppressed parametrically by
\begin{equation}
(2\pi)^3 \left(\frac{k_{mr}^4}{k_{NL}^4}\right) (\log(k/k_{mr}))^2\,, \qquad 10^{5}\tilde{c}_{comb,0}^2 \frac{k^2 k_{NL,0}^{2}}{k_{NL}^4}\,.
\end{equation}
Note that in terms of the expansion in powers of $k^2/k_{NL}^2$, the $\tilde{c}^{2}_{comb}$ terms enter as corrections to $\pi_2$, however, due to the time dependence and the absence of loop integrals their effects on the power spectrum are comparable to the gravitational collapse terms in $\pi_3$.  Since the EFT terms arise due to integrating out small-scale degrees of freedom, we can consider them to be part of the one-loop effects when these additional degrees of freedom are taken into account.

\subsection{Stochastic terms}

As discussed in \cite{Baumann:2010tm, Carrasco:2012cv} (see also \cite{Pimentel:2012tw, Carrasco:2013mua}), integrating out the small-scale degrees of freedom can also generate a stochastic background for the $\pi$ field evolution.  In the UV theory this originates from couplings of the form
\begin{equation}\nonumber
\delta S = M_P^{2} \int d\tau d^{3}x \,a^2 \Bigg[(\nabla_{i}\nabla_{j}\pi) s^{ij}\Bigg]\,,
\end{equation} 
where $s^{ij}(\vec{x}, \tau)$ is a function of the small-scale degrees of freedom, including the $\pi$ field on small scales as well as vorticity and velocity dispersion after shell crossing and virialization.  This term will contribute a stochastic source in the equation of motion for $\Delta \pi_2$ on long scales, and while this will not alter the expectation value of the $\pi$ field, it can contribute to the power spectrum.

From the point of view of the $\pi$ field EFT on large scales, the small-scale degrees of freedom have been integrated out and therefore do not alter the action, but the effect of the stochastic terms can show up as a modification to the background state and therefore to the power spectrum.  We can estimate the size of this contribution as a Poisson noise term with a normalization given by the background $\pi$ power spectrum and voxels on the scale of $k_{NL}$, and therefore the stochastic correction to the power spectrum is on the order of
\begin{equation}
\frac{\Delta P_{stochastic}(\vec{k}, \tau)}{P_{0}(\vec{k}, \tau)} = \left(\frac{k}{k_{NL}}\right)^{3}\,.
\end{equation}
Comparing this to the EFT terms, this can be thought of as in between the one-loop and two-loop orders in size.

\subsection{Non-Gaussian contributions}

So far we have assumed that the initial conditions for the pion field fluctuations are Gaussian, and that the pion field bispectrum
\begin{equation}
\langle \tilde{\pi}(\vec{k}_1)\tilde{\pi}(\vec{k}_2)\tilde{\pi}(\vec{k}_3)\rangle = (2\pi)^{3}B(\vec{k}_1, \vec{k}_2, \vec{k}_3) \delta(\vec{k}_1 + \vec{k}_2 + \vec{k}_3)
\end{equation}
is therefore vanishing at early times.  There may, however, be primordial non-Gaussianities arising from inflation, which can be parameterized by
\begin{equation}
\frac{\langle \Phi \Phi \Phi \rangle'}{P_{\Phi}^2} \sim f_{NL}\,,
\end{equation}
where the prime indicates that the momentum conserving delta function has been removed from the correlation function, and $P_{\Phi}$ is the power spectrum for the Newtonian potential $\Phi$.  
This gives a contribution to the reduced pion bispectrum $Q(\vec{k}_1, \vec{k}_2, \vec{k}_3) = B(\vec{k}_1, \vec{k}_2, \vec{k}_3)/(P(k_1)P(k_2) + P(k_2)P(k_3) + P(k_1)P(k_3))$ which is on the order of $\mathcal{H}f_{NL}$.

\begin{figure}[htbp]
\begin{center}
\includegraphics[scale=0.8]{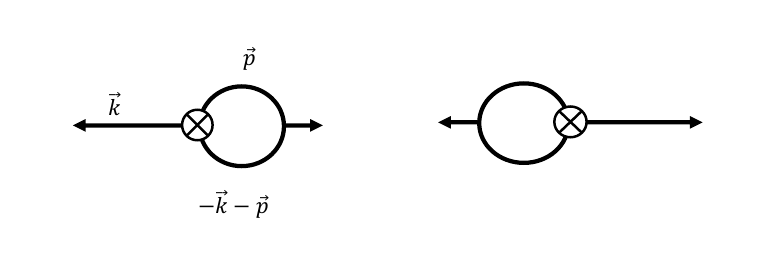}
\caption{One-loop contributions to the power spectrum from the bispectrum terms in the initial conditions.}
\label{Feynman_diagrams_NG}
\end{center}
\end{figure}

In Fig.~\ref{Feynman_diagrams_NG} we depict the diagrams that the initial conditions bispectrum contribute to the power spectrum at late times.
These give a corresponding contribution to the power spectrum which is equal to
\begin{equation}
\begin{split}
2\int \frac{d^{3}\vec{p}}{(2\pi)^3}&B(\vec{k}, \vec{p}, -\vec{k}-\vec{p})\int^{\tau}d\tau'\,G(\tau, \tau')\Bigg[(\mathcal{H}D^2 + 2DD')\vec{p}\cdot(-\vec{p}-\vec{k})\\
&+(\mathcal{H}D^2 + DD')\frac{1}{k^2}((\vec{p}\cdot\vec{k})(k+p)^2 - \vec{k}\cdot(\vec{k}+\vec{p})p^2)\Bigg]D(\tau)\,.
\end{split}
\end{equation}
To evaluate this precisely we would need the detailed form of the bispectrum; nevertheless, we can estimate the order of magnitude by estimating
\begin{equation}
B(\vec{k}, \vec{p}, -\vec{k}-\vec{p}) \sim \mathcal{H} f_{NL}\Big(P(k)P(p) + P(p)P(|\vec{k}+\vec{p}|) + P(|\vec{k}+\vec{p}|)P(k)\Big)
\end{equation}
and assuming that the dominant contributions to the integral are from $p \sim k$.  For primordial local-type non-Gaussianity with a value of $f_{NL}$, we therefore have
\begin{equation}\label{fNL_corrections_2PT}
\begin{split}
\frac{\Delta P_{NLO}(k)}{P_{0}(k)} &\sim 10^{-5}f_{NL}(2\pi)^3 \frac{k^{2}}{k_{NL}^{2}} \qquad \qquad \qquad \qquad \, \, \, \, \, \, k \lesssim k_{mr}\,,\\
&\sim 10^{-5}f_{NL}(2\pi)^3 \frac{k^{4}_{mr}}{k^2 k_{NL}^{2}}(\log(k/k_{mr}))^2 \qquad k \gtrsim k_{mr}\,,
\end{split}
\end{equation}

Comparing these to the corrections arising from the nonlinear evolution of $\pi$ due to gravitational collapse, we therefore expect that the contributions from primordial non-Gaussianity will be suppressed at late times.

\subsection{Consistency relations and the perturbative calculation}

For modes $k \gg k_{mr}$, if we approximate $p \ll k$ when performing the integrals in Eq.~\eqref{PT_LCDM_one}, Eq.~\eqref{PT_LCDM_two}, Eq.~\eqref{PT_LCDM_three}, then these simplify considerably to give the following result for the unequal-time correlator:
\begin{equation}\label{frac_Delta_one}
\begin{split}
\frac{\Delta P_{NLO}(k, \tau_1, \tau_2)}{P_{0}(k, \tau_1, \tau_2)} &= \frac{k^2}{6\pi^2}\int_{0}^{\infty}dp\, p^4 \hat{P}_{0}(p) \times \Bigg(\frac{D_{2}(\tau_1) D_{2}(\tau_2)}{D(\tau_1)D(\tau_2)}\\ 
&- \frac{D_{3}(\tau_1)}{D(\tau_1)} + \frac{D_{4}(\tau_1)}{D(\tau_1)} - \frac{D_{3}(\tau_2)}{D(\tau_2)} + \frac{D_{4}(\tau_2)}{D(\tau_2)}\Bigg)\,,\\
\end{split}
\end{equation}
where $D_{2} = D_{2a} + D_{2b}$, and $D_{3} = D_{3a} + D_{3b}$.  This contribution, proportional to $k^2$, vanishes when $\tau_1 = \tau_2$, and can be shown to be a consequence of the consistency relations for the spontaneously broken diffeomorphism symmetry given in Eq.~\eqref{EFTsymms}.

Consider the value of an N-point correlation function $\langle \mathcal{O}_{\vec{k}_1}\cdots \mathcal{O}_{\vec{k}_N}\rangle_{\pi_{soft}}$ in the presence of a background $\pi_{soft}$ of soft (low-momentum) modes at initial time $\tau_0$.  Taking the Taylor expansion in the soft momentum and averaging over a statistical ensemble of modes, we can expand
\begin{equation}
\begin{split}
\langle\langle \mathcal{O}_{\vec{k}_1}\cdots \mathcal{O}_{\vec{k}_N}\rangle_{\pi_{soft}}\rangle &\approx \langle \mathcal{O}_{\vec{k}_1}\cdots \mathcal{O}_{\vec{k}_N}\rangle_{0} + \int_{\vec{p}}\frac{\delta \langle \mathcal{O}_{\vec{k}_1}\cdots \mathcal{O}_{\vec{k}_N}\rangle_{\pi_{soft}}}{\delta \pi^{*}_{\vec{p}}}\bigg|_{0} \langle \pi_{\vec{p}} \rangle\\
&+ \frac{1}{2}\int_{\vec{p}_1, \vec{p}_2}\frac{\delta \langle \mathcal{O}_{\vec{k}_1}\cdots \mathcal{O}_{\vec{k}_N}\rangle_{\pi_{soft}}}{\delta \pi^{*}_{\vec{p}_1}\delta\pi^{*}_{\vec{p}_2}}\bigg|_{0} \langle \pi_{\vec{p}_1}\pi_{\vec{p}_2}\rangle + \cdots
\end{split}
\end{equation}
The one-point function vanishes, and the term with the two-point function becomes
\begin{equation}\label{Npt_shifted}
\langle\langle \mathcal{O}_{\vec{k}_1}\cdots \mathcal{O}_{\vec{k}_N}\rangle_{\pi_{soft}}\rangle \approx \langle \mathcal{O}_{\vec{k}_1}\cdots \mathcal{O}_{\vec{k}_N}\rangle_{0} + \frac{1}{2}\int_{\vec{p}}\frac{\delta \langle \mathcal{O}_{\vec{k}_1}\cdots \mathcal{O}_{\vec{k}_N}\rangle_{\pi_{soft}}}{\delta \pi^{*}_{\vec{p}}\delta\pi^{*}_{-\vec{p}}}\bigg|_{0} P(p, \tau_0)\,.
\end{equation}
Since the $\pi$ field is a Goldstone boson of the spontaneously broken symmetry, in the soft momentum limit the field can be removed by a symmetry transformation.  In the notation of \cite{Horn:2014rta}, we have
\begin{equation}
\int_{\vec{p}}\frac{\delta \langle \mathcal{O}_{\vec{k}_1}\cdots \mathcal{O}_{\vec{k}_N}\rangle_{\pi_{soft}}}{\delta \pi^{*}_{\vec{p}}}\bigg|_{0}\Delta_{nl}\pi_{\vec{p}} = \Delta_{lin}\langle \mathcal{O}_{\vec{k}_1}\cdots \mathcal{O}_{\vec{k}_N}\rangle\,,
\end{equation}
where $\Delta_{lin}, \Delta_{nl}$ are the parts of the transformation $\vec{x} \to \vec{x} + \vec{n}$, we have $\pi \to \pi - \vec{n}\cdot \vec{\nabla}\pi + \vec{n}'\cdot \vec{x}$ which are linear and nonlinear in the $\pi$ field, respectively.  We further require the adiabatic mode condition that $\vec{n} \propto \int D(\tau) d\tau$ in order for the transformed field to be a solution to the linearized equations of motion.  Putting this together, we have
\begin{equation}
\int_{\vec{p}}\frac{\delta \langle \mathcal{O}_{\vec{k}_1}\cdots \mathcal{O}_{\vec{k}_N}\rangle_{\pi_{soft}}}{\delta \pi^{*}_{\vec{p}}}\bigg|_{0}i \vec{n}'(\tau_0)\cdot \vec{\nabla}_{\vec{p}}\Big((2\pi)^{3}\delta^{(3)}(\vec{p})\Big) = \sum_{a} i \vec{n}(\tau_a) \cdot \vec{k}_{a}\langle \mathcal{O}_{\vec{k}_1}\cdots \mathcal{O}_{\vec{k}_N}\rangle\,,
\end{equation}
and therefore
\begin{equation}
\frac{\delta \langle \mathcal{O}_{\vec{k}_1}\cdots \mathcal{O}_{\vec{k}_N}\rangle}{\delta \pi^{*}_{\vec{p}}}\bigg|_{0} = -\sum_{a}\frac{D_0(\tau_a)}{D(\tau_0)}(\vec{p} \cdot \vec{k}_a)\langle \mathcal{O}_{\vec{k}_1}\cdots \mathcal{O}_{\vec{k}_N}\rangle\,,
\end{equation}
where we have written $D_0(\tau) = \int D(\tau') d\tau'$.  This is the consistency relation of \cite{Kehagias:2013yd, Peloso:2013spa} applied to the $\pi$ field formalism.  Applying this to \eqref{Npt_shifted}, we have
\begin{equation}\label{Npt_shifted}
\begin{split}
\langle\langle \mathcal{O}_{\vec{k}_1}\cdots \mathcal{O}_{\vec{k}_N}\rangle_{\pi}\rangle &\approx \langle \mathcal{O}_{\vec{k}_1}\cdots \mathcal{O}_{\vec{k}_N}\rangle_{0} \\ + \frac{1}{2}\sum_{a,b}\frac{D_0(\tau_a)D_0(\tau_b)}{D(\tau_0)^2}&\int_{\vec{p}} (\vec{p} \cdot \vec{k}_a)(\vec{p} \cdot \vec{k}_b) P(p, \tau_0) \langle \mathcal{O}_{\vec{k}_1}\cdots \mathcal{O}_{\vec{k}_N}\rangle\,.
\end{split}
\end{equation}
For the special case of the unequal-time two point function for $\pi$, we have the fractional correction
\begin{equation}\label{frac_Delta_two}
\begin{split}
\frac{\Delta P_{NLO}(k, \tau_1, \tau_2)}{P_{0}(k, \tau_1, \tau_2)} &= -\frac{1}{2}\int \frac{d^{3}\vec{p}}{(2\pi)^3}(\vec{p}\cdot \vec{k})^2 \hat{P}_{0}(p)\big(D_0(\tau_1)-D_0(\tau_2)\big)^{2}\\
&= -\frac{k^2}{12\pi^2}\int dp\, p^4 \hat{P}_{0}(p)\big(D_0(\tau_1)-D_0(\tau_2)\big)^{2}\,.
\end{split}
\end{equation}
Comparing Eq.~\eqref{frac_Delta_one} and Eq.~\eqref{frac_Delta_two}, we note that
\begin{equation}
D_{2}(\tau) = \int^{\tau} d\tau'\, G(\tau, \tau') (2 \mathcal{H}D^2 + 3DD')
\end{equation}
obeys $D_{2}' + 2\mathcal{H}D_{2} = \int^{\tau} d\tau'\, (2 \mathcal{H}D^2 + 3DD')$, and using $D' + 2\mathcal{H}D = 5$ and Eq.~\eqref{Green_func}, we have
\begin{equation}
\begin{split}
\Bigg(\frac{D_{2}}{D}\Bigg)' &= \frac{1}{D}\int d\tau'\,(2 \mathcal{H}D^2 + 3DD') - \frac{5}{D^2}\int d\tau' \, G(\tau, \tau') (2\mathcal{H}D^2 + 3DD')\\
&= \frac{1}{D^2 a^2}\int^{\tau} d\tau' a^2(2\mathcal{H}D^3 + 3D^2 D') = \frac{1}{D^2 a^2} \int^{\tau} d\tau' \, (a^2 D^3)' = D(\tau)\,,
\end{split}
\end{equation}
and therefore $D_2 = D_{0}D$.  We can also write
\begin{equation}
(D_{3}(\tau) - D_4(\tau)) = \int d\tau' G(\tau,\tau') (2\mathcal{H}D D_2 + D'D_2 + 2DD'_2 - D^3)\,,
\end{equation}
where $D_{2} = D_{2a} +  D_{2b}$.  Using a similar process as before, together with the identity $D_{2}' = D^2 + D_2 D'/D$, we have
\begin{equation}
\begin{split}
\Bigg(\frac{(D_{3}-D_4)}{D}\Bigg)' &= \frac{1}{D}\int d\tau' (2\mathcal{H}D D_2 + D'D_2 + 2DD'_2 - D^3) \\
&\qquad - \frac{5}{D^2}\int d\tau' \, G(\tau, \tau') (2\mathcal{H}D D_2 + D'D_2 + 2DD'_2 - D^3)\\
&= \frac{1}{D^2 a^2}\int^{\tau} d\tau' a^2(2\mathcal{H}D^2 D_2 + DD' D_2 + 2 D^2 D'_2 - D^4) \\
&= \frac{1}{D^2 a^2}\int^{\tau} d\tau' a^2(2\mathcal{H}D^2 D_2 + 2DD' D_2 + D^2 D'_2)\\
 &= \frac{1}{D^2 a^2}\int^{\tau}d\tau' (a^2 D^2 D_2)' = D_2 = D_0 D \,,
\end{split}
\end{equation}
and so $(D_3 - D_4) = \frac{1}{2}D_{0}^2$.  This completes the proof that Eq.~\eqref{frac_Delta_one} is equivalent to Eq.~\eqref{frac_Delta_two}.

The vanishing of the correction to $\Delta P_{NLO}(k, \tau)/P_{0}(k, \tau)$ that is quadratic in $k$, when $k \gg k_{mr}$, is therefore guaranteed by the consistency relations.  However, as discussed in detail in \cite{Horn:2014rta}, the consistency relations may be violated by the EFT terms on small scales, which change the time-dependent evolution of the linear modes, or by certain types of non-Gaussian corrections in the initial state.  As shown in Eq.~\eqref{cs2_corrections_2pt}, Eq.~\eqref{fNL_corrections_2PT}, both these statements are borne out by the perturbative calculations of the previous sections.

\section{Numerical simulation of the pion field}

In this section we numerically simulate the evolution of the $\pi$ field over cosmic time in order to visualize the field evolution and to compare the nonlinear evolution of the power spectrum to the perturbative calculations of the previous section.  We perform the evolution on a discrete spatial grid and with discrete (conformal) time steps.  The full 3D numerical simulation code is available on GitHub and can be used in Python or in C++.  We start with a simple toy model in 1D for simplicity, and then develop the full 3D version.
 

The full equation of motion \eqref{pi_eom_corrected} is a nonlocal integrodifferential equation because of the inverted Laplacian operator.  In 1D, the inverse of a derivative is simply an integral, so the operator $\nabla^{-2}\vec{\nabla}$ becomes an integral, and so the inverse of the Laplacian operator can be simply realized on a periodic grid from $x = 0$ to $x = L$ as the integral
\begin{equation}
\begin{split}
\frac{1}{\nabla^2}\nabla_i &\left(\nabla_i \pi \nabla^2 \Bigg[\pi' + \mathcal{H}\pi + \frac{1}{2}(\nabla \pi)^2 \Bigg]\right) = \int^{L}_{0}dx' \left(\nabla_i \pi \nabla^2 \Bigg[\pi' + \mathcal{H}\pi + \frac{1}{2}(\nabla \pi)^2 \Bigg]\right)\\ &- \frac{x}{L}\int^{L}_{0}\left(\nabla_i \pi \nabla^2 \Bigg[\pi' + \mathcal{H}\pi + \frac{1}{2}(\nabla \pi)^2 \Bigg]\right)\,.
\end{split}
\end{equation}
The runtime for this method scales like $O(N)$ in the number of spatial grid points $N$, and the second term, which is linear in $x$ and has vanishing Laplacian, is necessary to ensure that the result of inverting the Laplacian is a periodic function in $x$.  In 3D the inverse of the Laplacian operator can be realized in several different ways.  One method is to implement this using the Green's function  
\begin{equation}
G(\vec{r}, \vec{r}_0) = \sum_{\vec{n}}\int \frac{d^{3}\vec{r}'}{4 \pi |\vec{r} - \vec{r}_0 - \vec{r}' + \vec{n}|}\,, \qquad \nabla^2 G = \delta^{3}(\vec{r}-\vec{r}_0)\,,
\end{equation}
where for periodic boundary conditions the sum is taken over all lattice points.  The sum is conditionally convergent and can be evaluated using the Ewald formula \cite{Ewald:1921}.  However, numerically evaluating the Green's function uses on the order of $O(N^6)$ steps, since the integral must be performed once for each lattice point.  Furthermore, the finite lattice spacing means that the Green's function $1/(4\pi|\vec{r}-\vec{r}_0|)$ will be cut off, and will not evaluate to a perfect delta function when the finite difference Laplacian is taken.

A more efficient method to invert the Laplacian operator is to make use of the Fourier transform
\begin{equation}
\frac{1}{\nabla^2}f(\vec{x}) = \int \frac{d^3 \vec{k}}{(2\pi)^3}\frac{\tilde{f}(\vec{k})}{-\vec{k}^2 + \epsilon^2}e^{-i \vec{k}\cdot\vec{x}}\,, \qquad \textrm{where}\qquad \tilde{f}(\vec{k}) = \int d^{3}\vec{x} f(\vec{x}) e^{i \vec{k}\cdot\vec{x}}\,.
\end{equation}
Discretizing and using the Fast Fourier Transform algorithm reduces the scaling of the algorithmic complexity to $O(N^3 \log N)$ instead of $O(N^6)$.  To avoid the singularity at $\vec{k} = 0$, we introduce a small cutoff value $\epsilon$ into the denominator.

The discrete spatial lattice and time steps will introduce errors suppressed by powers of
\begin{equation}
k \Delta x\,, \qquad k/k_{Nyquist}\,, \qquad \mathcal{H}\Delta \tau\,,
\end{equation}
for the space and time derivatives respectively.  These need to be kept parametrically small in order to achieve accurate results.

\subsection{Shock formation in 1D}

We first simulate the evolution of the $\pi$ field in 1D using Eq.~\eqref{pi_eom}.  The spatial coordinate $x$ is discretized into $N$ evenly spaced lattice points, and in the $\Lambda$CDM universe, we find it more convenient to use the Friedman equations to convert the conformal time steps to steps in the scale factor.  Spatial derivatives and integrals are replaced by their finite difference counterparts
\begin{equation}
f'(x_i) = \frac{f(x_{i+1})-f(x_{i-1})}{2\Delta x}\,, \qquad \int f(x) dx = \sum_{i} f(x_i) dx\,,
\end{equation}
The index $i$ is counted modulo the number of lattice points, in order to implement periodic boundary conditions.  Because the integral can be stepped from one lattice point to the next, it is not necessary to recalculate the integral at each lattice point, and therefore the runtime scales as $O(N)$.

\begin{figure}
\begin{center}
\includegraphics[scale=0.4]{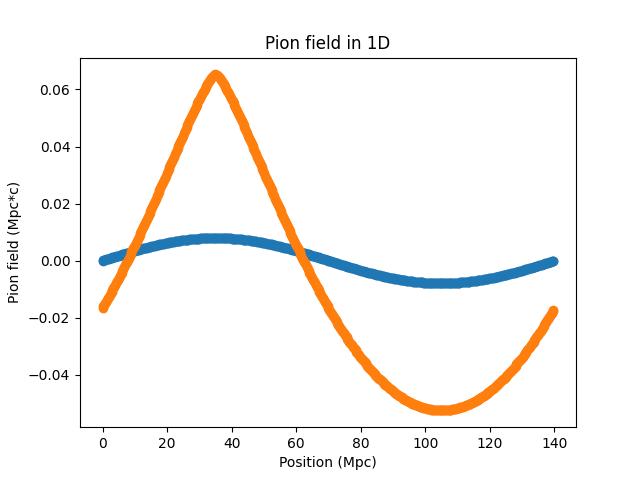}
\includegraphics[scale=0.4]{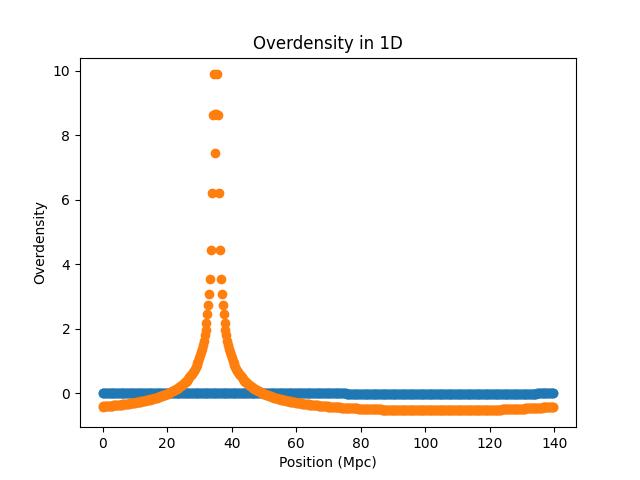}
\caption{Numerical simulation of $\pi(x)$ and $\delta(x)$ up to the time of shock formation, starting from sine-wave initial conditions.  Blue\: a = 0.01, Orange\: a = 1.\label{pi_1d_slice}}
\end{center}
\end{figure}

For our 1D warm-up, we consider a periodic spatial box with $L$ = 140 Mpc and $N = 500$ grid points in the benchmark universe with $\Omega_{m0} = 0.31$ and $H_0 = 68 km/s/Mpc$.  We start with $a = 0.01$ and the sine-wave initial conditions
\begin{equation}
\pi(x) = \frac{10^{-5}c}{\mathcal{H}}\sin(2\pi x/10 \textrm{Mpc})\,,
\end{equation}
and we keep the EFT coefficients $c_{s}^{2}, c_{v}^{2}$ set to zero.  The results of the numerical simulation for the evolution show the formation of a shock around $a = 1$, which is depicted in Fig.~\ref{pi_1d_slice}.  
Note that since $\vec{\nabla}\pi = \vec{v}$, the part of the curve where the velocity transitions from positive to negative signals the formation of a shock wave, beyond with the single-stream approximation will break down.  We can estimate that this occurs when
\begin{equation}
v \sim \frac{\Delta \pi}{\Delta x} \sim \frac{\Delta x}{\Delta \tau} \rightarrow \Delta \pi \sim \frac{(\Delta x)^2}{\Delta \tau}\,.
\end{equation}
This is also consistent with the effective field theory power counting in \S 3.1, which predicts that nonlinearities occur when $k \sim k_{NL}$.  For this example, we can check numerically that
\begin{equation}
\pi \sim \frac{\mathcal{H} c}{k^2} \sim \frac{c/(4408\textrm{Mpc})}{((2\pi)/140\textrm{Mpc})^2} \sim (0.1)\textrm{Mpc}  \cdot c\,.
\end{equation}
Plotting the corresponding $\delta$ field, we see that the finite difference method becomes less accurate for highly collapsed structures, and continuing the simulation beyond this point leads to the formation of unphysical regions where $\delta < -1$ around the edges of the halos, and the $\pi$ field will become highly oscillatory and diverge.  At this point additional degrees of freedom will be required in the theory and the $\pi$ field picture breaks down.

\subsection{3D simulation: PLASTIC}

To perform the numerical simulation in three dimensions, we discretize space into a grid of $N^3$ lattice points.  Both the grid points and the $\pi$ field data can be implemented as nested arrays, and the evolution proceeds in steps of discrete conformal time $\Delta \tau$.  As before, in the $\Lambda$CDM universe, we find it more convenient to use the Friedman equations to convert this to steps in the scale factor.  At each step, the evolution of the $\pi$ field is realized using the present data for $\pi$ and $\pi'$, and using the main equation of motion Eq.~\eqref{pi_eom_corrected}.  We use Fourier transforms to implement the spatial derivatives.


The full code is available in both Python and C++ on GitHub\footnote{https://github.com/bhorn9/pion\_lss\_code} under the name Pion LAgrangian for STructure In Cosmology (PLASTIC) and contains the following programs:
\begin{itemize}
\item GaussianIC: generates Gaussian initial conditions for the $\pi$ field with a desired power spectrum.
\item PLASTIC(++): evolves the $\pi$ field over cosmic time (can be implemented in Python or C++). 
\item Plots: graphs the power spectra and fields in 3D.
\end{itemize}
The advantages of Python include ease of coding and modification, and the use of existing libraries such as numpy for scientific computation and matplotlib for graphics.  We find a runtime of approximately 1-2 seconds per time step for $N = 50$, which represents a factor of around 20 times improvement on the runtime using Wolfram Mathematica.  Using C++ instead considerably improves the runtime by a further factor of 20 or so over Python, but requires either the installation of external libraries to implement the Fast Fourier Transform \cite{FFTW} or writing our own FFT program.  For the latter option, we have made use of the Cooley-Tukey\cite{Cooley:1965} algorithm to implement the Fast Fourier Transform in C++ for values of $N$ that have only 2, 3, and 5 as prime factors.  The output can be written as a text file, and then plotted using Python.  The runtime achieved is comparable to using numpy but still slower than FFTW.

Gaussian initial conditions can be generated in Fourier space by setting the phase to be random to and setting the amplitude to match the desired power spectrum.  We can use either the fitting function of \cite{Eisenstein:1997ik} or \cite{Bardeen:1985tr} to estimate the power spectrum, or fit to a given data file.  We generate the linear power spectrum for a benchmark $\Lambda$CDM universe with $\Omega_{m0} = 0.31$, $H_0 = 68$km/s/Mpc using the power spectrum data generated by the Cosmic Linear Anisotropy System Solver (CLASS) \cite{Blas:2011rf}.
Evolving the pion field over time, we expect that shock waves will form when the field enters the nonlinear regime at around
\begin{equation}
\pi \sim \frac{\mathcal{H}}{\vec{k}^2}\,,
\end{equation}
which also corresponds to the regime where the overdensity $\delta \sim 1$.  In keeping with the lessons learned from our 1D simulation, we terminate the simulation if $\delta$ is about to dip below -1, which can occur due to numerical errors introduced by the finite size of the lattice in regions where nearly empty space transitions suddenly to a compact halo. 

\begin{figure}
\begin{minipage}{.5\textwidth}
\includegraphics[scale=0.6]{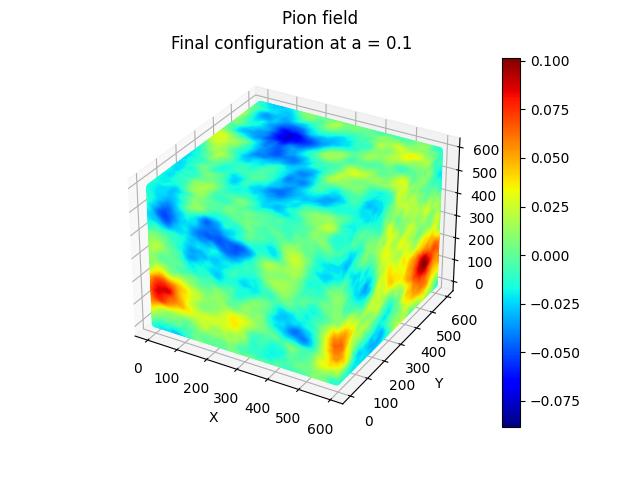}
\end{minipage}
\begin{minipage}{.5\textwidth}
\includegraphics[scale=0.6]{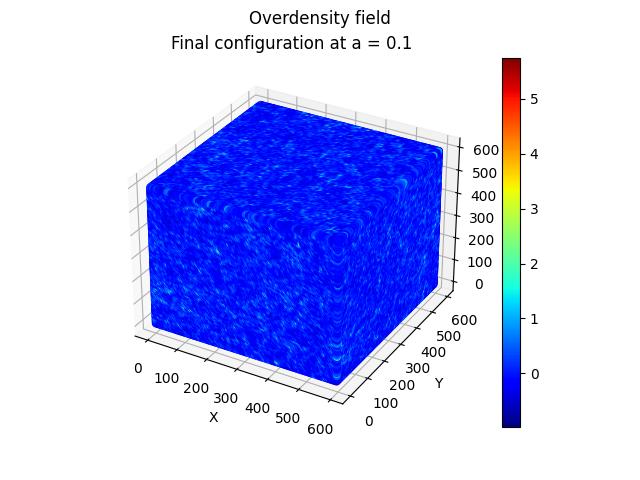}
\end{minipage}
\caption{Evolved conditions for $\pi$ and $\delta$ at a = 0.1.  Boxsize = 600 Mpc, N = 128, $\Omega_m = 0.31$ and $H_0 = 6$8 km/s/Mpc.}
\label{output_3D}
\end{figure}

\begin{figure}
\begin{center}
\includegraphics[scale=0.45]{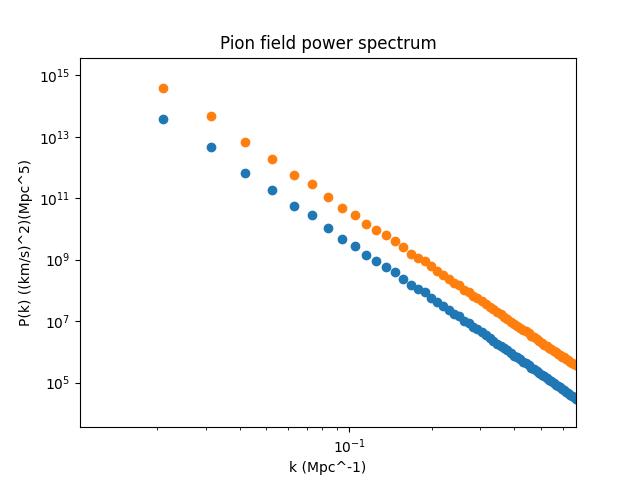}
\includegraphics[scale=0.45]{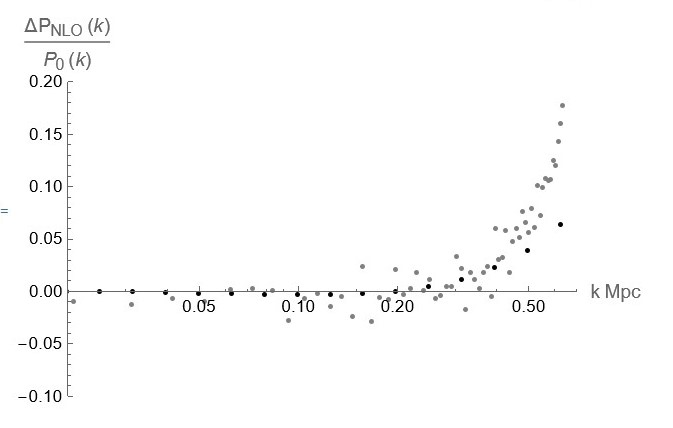}
\caption{Left: The evolution of the pion field power spectrum with box size = 600 Mpc, N = 128, $\Omega_{m0} = 0.31$, for a  = 0.01 (blue) and a = 0.10 (orange).  Right: Fractional corrections to the power spectrum from nonlinear collapse.  Black = analytic prediction at NLO, Gray = numerical simulation.}
\label{pi_power_spectrum}
\end{center}
\end{figure}

As a demonstration of the code, we start from initial conditions in the linear regime with $a = 0.01$, $\Omega_{m0} = 0.31$, a box size of 600 Mpc, and $N = 128$ grid points, corresponding to a Nyquist scale of $9.375$ Mpc.  The evolved configuration for the $\pi$ and $\delta$ field for the simulation parameters given above is shown in Fig.~\ref{output_3D}.  For the given choice of the box size and Nyquist scales, shock formation occurs very shortly after $a = 0.10$.
The power spectrum is plotted in Fig~\ref{pi_power_spectrum}.  While the power spectrum is observed to increase linearly on large scales, the simulated linear growth factor is consistently a few percent higher than the analytic prediction, due to numerical errors from the discrete time stepping.  Removing the constant offset this generates, the fractional change in the power spectrum due to the nonlinearities agrees reasonably well between the analytic prediction and numerical simulation.  

As stated above, the discrete time stepping will introduce an overall error on the order of $\mathcal{H}\Delta \tau$, and the discrete spatial grid will correct the derivatives by additional powers of $k/k_{Nyquist}$.  Indeed, we observe that for $k \gtrsim k_{Nyquist}$ the numerical results deviate from the NLO perturbative calculation.  Higher order nonlinear terms are suppressed by powers of $k/k_{NL}$, and therefore only the leading order nonlinear effects are visible with accuracy at this grid resolution.  


\section{Comparison with N-body simulations}

To further understand the evolution of large scale structure in the pion field picture, we analyze the results of an N-body simulation.  We make use of the software GADGET-4 \cite{Springel:2020plp} to simulate the evolution of $128^{3} = 2,097,152$ individual cold dark matter particles in a $\Lambda$CDM universe with $\Omega_m = 0.308$ and $H_0 = 67.8$km/s/Mpc, for a periodic box with size $400$ Mpc/h $\approx 590$ Mpc and a particle-mesh grid of $384^3$ grid points.  The initial conditions are evolved from redshift $z = 63$ to the present day.

\begin{figure}
\begin{center}
\includegraphics[scale=0.7]{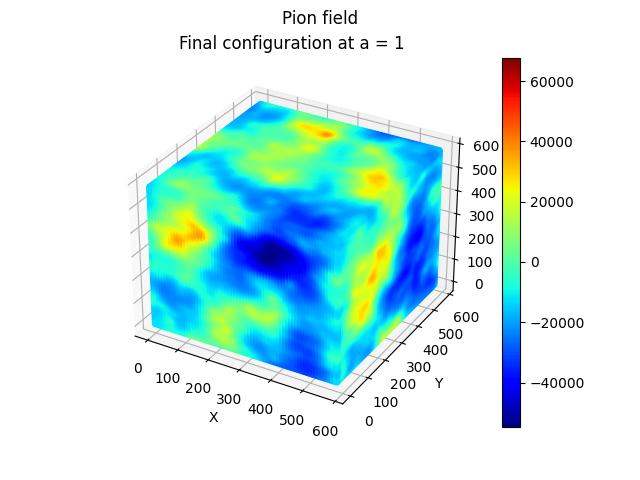}
\caption{The pion field at $z=0$ from an N-body simulation with box size 590 Mpc, $\Omega_{m0} = 0.308$, $H_0 = 67.8$km/s/Mpc and $2^{21}$ particles.  The pion field is measured in units of Mpc(km/s).}
\label{pi_simulated}
\end{center}
\end{figure}

We make use of the PyLiANS libraries \cite{Pylians} to construct smoothed fields from the simulation output.  We make use of the additivity of mass and momentum to calculate $\rho$, $\rho \vec{v}$, within each voxel of a 3D grid, optionally smoothing with a Gaussian filter.  We then calculate $\delta$ and $\vec{v}$, and use the Fourier transform to calculate the $\pi$ field as a function of position.  For further discussion of the subtleties of reconstructing the cosmic velocity field in the multi-stream regime, see e.g. \cite{Hahn:2014lca}.  For a grid of $64^3$ voxels, a snapshot of the pion field at $z = 0$ is shown in Fig.~\ref{pi_simulated}.

\subsection{Power spectrum}
  
\begin{figure}
\begin{center}
\includegraphics[scale=0.7]{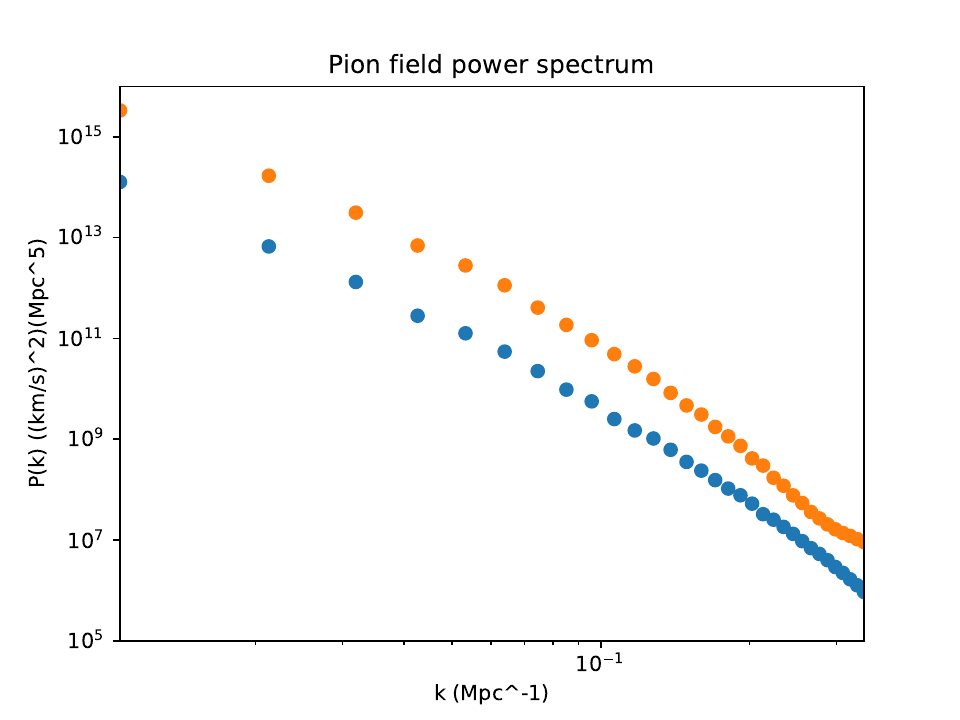}
\caption{The pion field power spectrum at $z=0$ from an N-body simulation with box size $400/h$ Mpc, $h = 0.678$, $\Omega_{m0} = 0.308$ and $2^{21}$ particles.  Blue: z = 49, Orange: z = 0.}
\label{pi_Pk_simulated}
\end{center}
\end{figure}

\begin{figure}
\begin{center}
\includegraphics[scale=0.5]{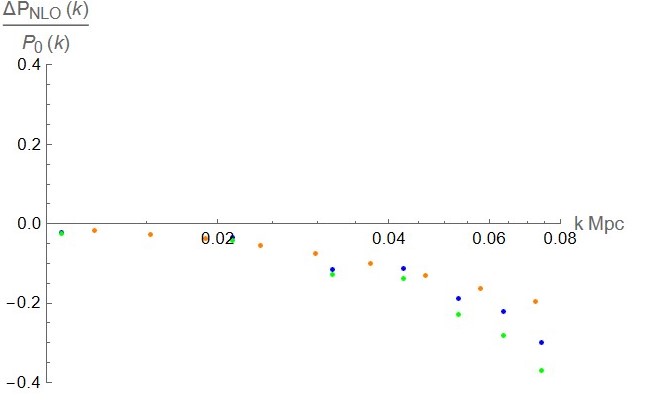}
\caption{The fractional corrections to the pion field power spectrum from an N-body simulation with box size $400/h \approx 590$ Mpc with $h = 0.678$, $\Omega_{m0} = 0.308$ and $2^{21}$ particles.  Blue = measured nonlinear power spectrum for a grid of $64^3$ voxels, Green = grid resolution increased to $128^3$ voxels and a Gaussian filter of 10Mpc applied, Orange = predicted power spectrum from leading order in perturbation theory.}
\label{pi_Pk_NLO_simulated}
\end{center}
\end{figure}

The pion field power spectrum for our simulation at $z = 0$ is shown in Fig.~\ref{pi_Pk_simulated}, and the nonlinear corrections are plotted in Fig.~\ref{pi_Pk_NLO_simulated}.  The exact values for the $\pi$ field depend on the smoothing procedure used, so we compare the results of smoothing $128^3$ particles on a grid of $64^3$ voxels versus using a Gaussian filter of similar size.

The power spectrum roughly agrees with the perturbation theory calculation in \S 2 up to a scale of $k \sim 0.05$/Mpc, with slight deviations due to cosmic variance and the binning procedure used to construct the histogram.  At larger values of $k$, the N-body calculations include corrections in powers of $k/k_{NL}$ coming from higher-order perturbation theory as well as stochastic terms \cite{Baumann:2010tm}.  At higher values of $k$, the simulated power spectrum will receive additional corrections suppressed by powers of $k/k_{Nyquist}$ and $k/\Lambda$, where $2\pi/\Lambda$ is the smoothing scale in position space.



\begin{figure}
\begin{center}
\includegraphics[scale=0.6]{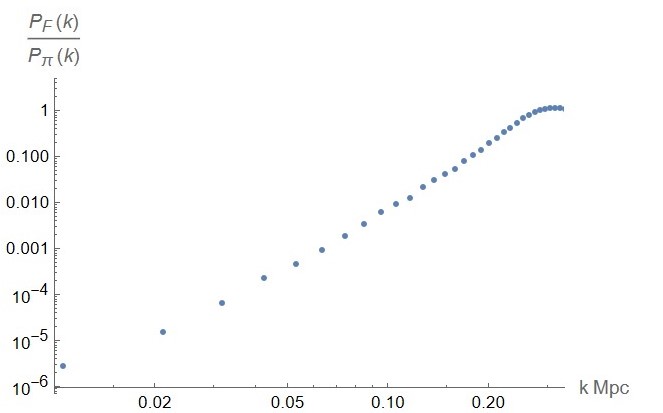}
\caption{The ratio of the power spectra for the velocity vector potential and the velocity scalar potential, at $z = 0$ for an N-body simulation with box size $400/h \approx 590$ Mpc, $\Omega_{m0} = 0.308$, $H_0 = 67.8$km/s/Mpc and $2^{21}$ particles.}
\label{pi_F_Pk_simulated}
\end{center}
\end{figure}

We can also measure the power spectrum for the velocity vector potential
\begin{equation}
\vec{v} = \vec{\nabla} \pi + \vec{\nabla} \times \vec{F}\,,
\end{equation}
which is depicted in Fig.~\ref{pi_F_Pk_simulated}.  As described in \S 2, the vector potential is small at large scales and early times, and is generated on small scales and late times due to shell crossing.  For $k \gtrsim k_{NL}$, the measured power is similar for the scalar and vector modes of the velocity, signaling the breakdown of the gradient flow approximation on these scales.  For $k \ll k_{NL}$, we find that at late times the power spectra obey $P_{F}(k)/P_{\pi}(k) \propto 1/k^{5}$.  This can be understood as follows: taking the curl of the Euler equation as in Eq.~\eqref{vorticity} and starting with $\omega \approx 0$, the velocity potential will be sourced by the vector potential of the stress tensor divergence $\nabla_{j}\tau^{ij}$, just as the equation of motion for $\pi$ has an EFT source term from the scalar potential.  If these terms were of similar size in the EFT expansion, we would have
\begin{equation}
F'^{i} \sim \left(\frac{c_s^{2}}{10^{-5}}\right)\frac{k^{i}}{k_{NL}^{2}}\pi\,.
\end{equation}
We also need to keep in mind, however, that the vorticity arises in nonlinear halos, with a direction in each that is independent of other such nonlinear structures.  There will therefore be a stochastic $(k/k_{NL})^{3/2}$ suppression for the vector potential modes.  Putting this together with the expression above gives the observed relative suppression for the velocity vector potential power spectrum.

\subsection{EFT parameters from simulation}

The sound speed and viscosity can be measured using the correlation functions, adapting the techniques of EFT for LSS in \cite{Baumann:2010tm, Carrasco:2012cv} to the $\pi$ field picture.  As discussed in \cite{Carrasco:2012cv, Pajer:2013jj}, the precise values will depend on the cutoff procedure used for coarse-graining the theory.  We can define
\begin{equation}
\mathcal{A} = \Phi + \pi' + \mathcal{H}\pi + \frac{1}{2}(\nabla \pi)^2\,,
\end{equation}
and then the sound speed and dimensionless viscosity parameters are given by
\begin{equation}\label{sound_speed_calc}
\begin{split}
c_{s}^2 &= \frac{3\mathcal{H}^2 \Omega_m}{2}\frac{\langle \mathcal{A} \pi' \rangle \langle (\nabla^2 \pi) \pi \rangle -\langle \mathcal{A} \pi \rangle \langle (\nabla^2 \pi) \pi' \rangle}{\langle (\nabla^2\pi) \pi \rangle \langle (\nabla^2\pi') \pi' \rangle - \langle (\nabla^2\pi) \pi' \rangle \langle(\nabla^2 \pi')\pi\rangle} \,, \\ \frac{2}{3\Omega_m}{c_{s}^2} &+ {c_{v}^{2}} = \mathcal{H}\frac{\langle \mathcal{A} \pi \rangle \langle (\nabla^2 \pi') \pi' \rangle - \langle \mathcal{A} \pi' \rangle \langle (\nabla^2 \pi') \pi \rangle}{\langle (\nabla^2\pi) \pi \rangle \langle (\nabla^2\pi') \pi' \rangle - \langle (\nabla^2\pi) \pi' \rangle \langle(\nabla^2 \pi')\pi\rangle}\,.
\end{split}
\end{equation}
We can calculate the two-point functions in position space, sampling over pairs of points to make a histogram and averaging in each distance bin.  We then take the average value over all bins for distances greater the smoothing scale and where both the numerator and denominators in Eq.~\eqref{sound_speed_calc} are measurably larger than zero above the level of random noise.

For our GADGET-4 simulation with $\Omega_{m0} = 0.308$, $H_0 = 67.8$km/s/Mpc, box size $400/h \approx 590$ Mpc, and a particle-mesh grid of $384^3$ points, we smooth the results at $z = 0$ with a Gaussian window function with scale $1/\Lambda$ to obtain smoothed fields and calculate the EFT coefficients.  The smoothing scales are chosen for comparison with \cite{Carrasco:2012cv} and the values obtained are given in the table below:

\begin{table}[h!]
\centering
\begin{tabular}{| c | c | c | c | c |}
 \hline
 $\Lambda (1/Mpc)$  & $10^{6}c_s^2$ & $10^{6}c_{vis}^{2} $ & $(4/3) \eta + \zeta $& $10^{6}\tilde{c}_{comb,init}^{2}$\\ 
 \hline
 $h/6$ & $(-1.2 \pm 0.2)$ & $(4.4 \pm 0.7)$ & $(4.8 \pm 0.8) \times 10^{5} cP$ & $(3.8 \pm 0.7) $ \\  
 $h/3$ & $(-0.4 \pm 0.1)$ & $(1.9 \pm 0.6)$ & $(2.1 \pm 0.7) \times 10^{5} cP$ & $(2.1 \pm 0.7) $\\
 \hline   
\end{tabular}
\caption{Mean values and estimated error bars for the squared sound speed and viscosity parameters at $z = 0$, using a GADGET-4 simulation with $\Omega_{m0} = 0.308$, $H_0 = 67.8$km/s/Mpc, and box size $400/h \approx 590$ Mpc.}
\label{EFT_params_table}
\end{table}
The sound speed squared is negative and the viscosity is positive, which means that nonlinear dynamics on small scales tends to cause instability and momentum diffusion on large scales.  In conventional units the viscosity parameter becomes $(4/3)\eta + \zeta = (\bar{\rho}_m c^2/\mathcal{H})c_{vis}^{2} = (1.1 \times 10^{11})c^2_{vis}$ centipoise (milliPascal-seconds), which yields a range for the viscosity of our simulated universe somewhere between peanut butter and Crisco \cite{vcc}.  The error bars given are estimates and include both a statistical component from calculating the correlation functions using the limited number of modes in the box, as well as a contribution on the order of $c^2 \Lambda \Delta x_{PM}$ due to numerical error from the finite size $\Delta x_{PM}$ of the particle-mesh grid.  
Both the numerical values and the approximate level of precision obtained using the GADGET-4 simulation are comparable to those obtained in \cite{Carrasco:2012cv} using a more powerful simulation with finer resolution, a larger number of particles, and a slightly different choice of $\Omega_{m0}$.

\begin{figure}
\begin{center}
\includegraphics[scale=0.5]{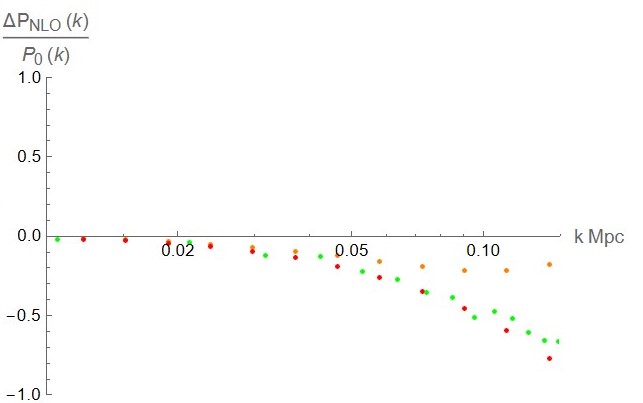}
\caption{The fractional corrections to the pion field power spectrum from an N-body simulation with box size $400/h \approx 590$ Mpc with $h = 0.678$, $\Omega_{m0} = 0.308$ and $2^{21}$ particles.  Orange = predicted power spectrum from leading order in perturbation theory.  Red = EFT terms included for $1/\Lambda = 6/h$Mpc.  Green = N-body simulation with Gaussian filter of $6/h$ Mpc applied}
\label{pi_Pk_NLO_simulated_GADGET}
\end{center}
\end{figure}

The values obtained from the numerical simulation can be used to estimate the scale of the EFT coefficients to be used in our top-down pion field simulation.  Using the ansatze $c_{s}^{2} \propto D^2, c_{v}^{2} \propto \Omega_{m}^{-1/2} D^2$  in Eq.~\eqref{tilded_EFT_coeffs}, we can find expressions for the tilded coefficients $\tilde{c}_{s}^{2}, \tilde{c}_{v}^{2}, \tilde{c}_{comb}^{2}$ as functions of time.   
For $z \gtrsim 1$ the time dependence is consistent within 10 percent with the approximations $\tilde{c}_{comb}^{2}(\tau) = \tilde{c}^{2}_{comb,init} (D(\tau)/D(\tau_0))^2$, where $\tilde{c}^{2}_{comb,init} = 3(c_{s0}^{2} + c_{v0}^{2}\sqrt{\Omega_{m0}})$ is given by $3.8 \times 10^{-6}$ for $\Lambda = h/6$ and $\tilde{c}_{comb}^{2}(\tau) = 2.1 \times 10^{-6}$ for $\Lambda = h/3$.  Returning to the fractional corrections to the pion field power spectrum depicted and including the EFT contributions in the perturbative calculations, we see in Fig.~\ref{pi_Pk_NLO_simulated_GADGET} that the fit is improved considerably out to $k \sim 0.15/$Mpc.


\section{Conclusion and further directions}

We have investigated the evolution of large scale structure in terms of the effective field theory of a single scalar pion field, where the $\pi$ field corresponds to the spontaneous breaking of spacetime symmetry in the presence of pressureless matter.  As discussed in \cite{Horn:2014rta}, this allows us to express the standard observables $\delta, \vec{v}, \Phi$ in terms of a single degree of freedom $\pi$, which can be described using a local action with an unusual derivative structure.  In the present work we have used this framework to investigate NLO corrections to the effective field theory in this picture, and we have used perturbation theory calculations and numerical simulations to study the evolution of the cosmic $pi$on field.  Writing the field theory in terms of $\pi$ keeps the size of the NLO corrections manifest, and allows us to rederive the NLO corrections to the evolution of the matter field and to the matter power spectrum in a systematic way.  Using this framework, we have compared results from perturbation theory against numerical simulations for the evolution of the pion field, and against N-body simulations to measure the EFT parameters for our simulation.  We find reasonable agreement between our analytical calculations and simulations in the weakly nonlinear regime, and we find that our perturbative calculations agree with the consistency relations described in \cite{Kehagias:2013yd, Peloso:2013spa, Horn:2014rta} in the appropriate limit.

Our effective field theory breaks down on small spatial scales due to shell crossing and virialization, which gives rise to vorticity, velocity dispersion, and an infinite tower of higher order moments in the Boltzmann hierarchy.  Corrections to the effective field theory that arise due to shell crossing on small scales can be absorbed into the $\pi$ field Lagrangian.  It would be interesting to explore ways of generalizing the pion field picture to include effects such as velocity dispersion and vorticity, either by using the fluid action derived in \cite{Endlich:2010hf} or by including additional degrees of freedom, perhaps along the lines of \cite{Pietroni:2018ebj, Erschfeld:2018zqg}.

We hope too that this picture will help motivate new observables to search for in surveys of large scale structure; as discussed in \cite{Horn:2014rta}, consistency relations are naturally expressed in terms of this variable.  It would be interesting to investigate further the realization of these consistency relations in simulations and in the experimental data, and we plan to return to this question in the future.  It is also of interest to further investigate the properties of $\pi$ field theory, given the unconventional structure of the kinetic terms, and we plan to explore this further as well.

\section*{Acknowledgments}

We thank Raphael Flauger, Simon Foreman, Lam Hui, Alberto Nicolis, David Spergel, Filippo Vernizzi, Zvonimir Vlah and Xiao Xiao for very helpful discussions, and we thank Lam Hui, Xiao Xiao and Eamon McShane for collaboration on related projects.  We thank the Office of the Dean of the Kakos School of Arts and Sciences and the Office of the Provost at Manhattan University, and we thank Kashifuddin Qazi, Jack Simoni and Katherine Uffer for assistance with computing resources at Manhattan University. This research is supported in part by the National Science Foundation under grant PHY-2210475, as part of the Research at Undergraduate Institutions program.




\bibliographystyle{JHEP}
\renewcommand{\refname}{Bibliography}
\addcontentsline{toc}{section}{Bibliography}
\providecommand{\href}[2]{#2}\begingroup\raggedright

\end{document}